\documentclass{ws-ijmpa}
\usepackage[super,compress]{cite}
\usepackage{graphicx}
\usepackage{color,amsmath,amsfonts,multirow}
\usepackage{epsfig}
\usepackage{rotating}
\usepackage{array}
\usepackage{graphicx}
\usepackage{multirow}
\usepackage{epstopdf}
\usepackage{amssymb}
\usepackage{latexsym}
\usepackage{comment}
\usepackage{bm}
\usepackage{accents}
\usepackage{tabularx}
\usepackage{booktabs}
\usepackage{mathtools}
\usepackage{slashed}
\usepackage{xspace}
\usepackage{slashed}
\usepackage{physics}
\usepackage{bbold}
\usepackage{esvect}
\usepackage{tikz}
\usepackage{tabularx}
\usepackage{booktabs}
\usepackage{xcolor}
\definecolor{darkblue}{rgb}{0.0, 0.0, 0.55}
\definecolor{lightbrown}{rgb}{0.71, 0.40, 0.11}
\usetikzlibrary{arrows.meta, positioning, shapes.misc, calc, fit, patterns}
\def\be{\begin{eqnarray}}
\def\ee{\end{eqnarray}}
\newcommand{\fig}{\begin{figure}}
\newcommand{\ef}{\end{figure}}

\newcommand{\nua}[1]{\ensuremath{\rlap{\kern-2.5pt\ensuremath{\overset{\scriptscriptstyle(-)}{\phantom{\nu}}}}{\ensuremath{{\nu}_{#1}}}}\xspace}

\newcommand{\MeV}{~\mathrm{MeV}}
\newcommand{\GeV}{~\mathrm{GeV}}

\newcommand{\ba}{\begin{align*}}
\newcommand{\ea}{\end{align*}}
\newcommand{\bpm}{\begin{pmatrix}}
\newcommand{\epm}{\end{pmatrix}}
\newcommand{\bea}{\begin{eqnarray}}
\newcommand{\eea}{\end{eqnarray}}

\newcommand{\cen}{CE$\nu$NS~}

\begin{document}

%
\catchline{}{}{}{}{}
%

\title{Neutrino Dipole Portal
}

\author{Sin Kyu Kang
}
\address{School of Natural Science, Seoul National University of Science Technology, Gongreung-ro 232\\
Seoul, 01811, Republic of Korea
\\
skkang@seoultech.ac.kr}

\author{C.J. Ouseph}
\address{Institute of Convergence Fundamental Studies, Seoul National University
of Science and Technology, Seoul 01811, Korea}

\maketitle

\begin{history}
\received{Day Month Year}
\revised{Day Month Year}
\end{history}

\begin{abstract}
The neutrino dipole portal (NDP) is a minimal and predictive extension of the Standard Model, in which a transition magnetic moment operator couples an active neutrino to a heavy neutral lepton via the electromagnetic field. 
This higher-dimensional interaction gives rise to distinctive processes such as neutrino up-scattering, radiative decays, meson transitions, and modifications of recoil spectra, offering multiple avenues for discovery. 
In this review, we discuss the theoretical foundations of the NDP, its ultraviolet completions, and the associated production and decay mechanisms across laboratory, astrophysical, and cosmological settings. 
Current constraints arise from accelerator searches, recoil-based detectors, collider studies, and high energy neutrino observatories, complemented by robust bounds from Big Bang Nucleosynthesis, the Cosmic Microwave Background, and supernova cooling. 
Future experimental and observational efforts, including next-generation neutrino experiments, multi-ton dark matter detectors, and improved cosmological and astrophysical probes, are anticipated to test the remaining allowed regions.
The NDP thus provides a simple, well-motivated, and broadly testable framework at the intersection of particle physics, astrophysics, and cosmology.
\end{abstract}




\section{Introduction}
\label{sec:intro}
Neutrino oscillation experiments~\cite{Super-Kamiokande:1998kpq,SNO:2002tuh,KamLAND:2002uet} provided unambiguous evidence for nonzero neutrino masses, a finding fundamentally incompatible with the Standard Model (SM), which admits only massless neutrinos.
This breakthrough transformed our view of fundamental interactions and motivated intensive theoretical and experimental research aimed at elucidating the origin of neutrino masses and their significance for physics beyond the SM.
A variety of mechanisms have been proposed to account for the smallness of neutrino masses, including canonical seesaw mechanisms~\cite{Minkowski:1977sc,Mohapatra:1979ia,Schechter:1980gr}, radiative generation models~\cite{Zee:1980ai,Ma:2006}, and extra-dimensional frameworks~\cite{Arkani-Hamed:1998wuz,Ma:2000wpa}. Despite impressive experimental progress, the precise mechanism remains elusive.

Among the many theoretical candidates, sterile neutrinos occupy a prominent place~\cite{Giunti:2019aiy,Kang:2019xuq,Dasgupta:2021ies}. Sterile neutrinos are electroweak singlets, insensitive to SM gauge interactions but potentially mixing with active neutrinos. They arise naturally in seesaw frameworks, where their inclusion can elegantly explain the lightness of active neutrino masses, and they frequently appear in models connecting the visible and hidden (dark) sectors~\cite{Escudero:2016tzx,Escudero:2016ksa,Ballett:2019pyw}. Their masses are essentially free parameters, spanning from the sub-eV scale to the Planck scale, although scenarios tied to neutrino mass generation often restrict them below the TeV scale. Because of this versatility, sterile neutrinos serve as natural extensions of the SM and as a bridge between laboratory anomalies, cosmological observations, and astrophysical phenomena~\cite{Kang:2019xuq,Boser:2019rta}.

Another important aspect of neutrino physics concerns their electromagnetic (EM) properties~\cite{Kim:2019add,Jeong:2021ivd,Kim:1976gk}.
In the SM, neutrinos are electrically neutral and possess only tiny magnetic moments, generated radiatively at higher order~\cite{Kim:1976gk,Fujikawa:1980yx,Giunti:2014ixa}. 
For Dirac neutrinos, the expected value is of order 
$\mu_\nu^{\text{SM}} \sim 10^{-19}(m_\nu/1\,\text{eV})\mu_B$, 
far below current experimental sensitivity~\cite{CONUS:2022qbb,ParticleDataGroup:2024cfk}. 
Historically, the idea of neutrino magnetic moments played a central role: Schwinger first speculated on their existence~\cite{schwinger}, Cisneros proposed them as a possible resolution of the solar neutrino problem~\cite{Cisneros:1970nq}, and many experiments have pursued them as clean signals of new physics.  

Beyond the SM, a wide range of theories can generate substantially larger magnetic or transition dipole moments~\cite{Giunti:2024gec}. In particular, if sterile neutrinos or heavy neutral leptons (HNLs) exist in the keV–GeV mass range, they can mix with active neutrinos and acquire transition dipole moments large enough to be observable. This possibility opens new avenues for discovery, allowing neutrinos to engage in EM interactions such as radiative decays or photon mediated scattering. Such processes are highly distinctive experimentally, often leading to clean single-photon signatures or distortions of recoil spectra at low energies.

A natural framework for describing these interactions is the \emph{Neutrino Dipole Portal} (NDP). The NDP is an effective field theory (EFT) extension in which a dimension-five operator couples an active neutrino to a HNL through the EM field tensor,
\[
\mathcal{L}_\text{NDP} \supset \mu_{\nu N} \, \bar{\nu}_L \sigma^{\mu\nu} N_R F_{\mu\nu},
\]
where $\mu_{\nu N}$ denotes the transition dipole moment. Hereafter, we denote $N\equiv N_R$ for simplicity and use the term "HNL" in place of sterile neutrino. This operator is chirality-flipping, gauge-invariant, and can be UV-completed in multiple ways, often involving heavy charged fermions or scalars in loops~\cite{Aparici:2009fh,Babu:2020ivd}. 
Its theoretical appeal lies in its minimality, as the entire phenomenology is governed by a single parameter $\mu_{\nu N}$, and in its robustness, since dipole operators naturally appear in effective Lagrangian once heavy fields are integrated out.

The phenomenology of the NDP is exceptionally rich, encompassing processes such as neutrino up-scattering $(\nu A \rightarrow N A,~ \nu e \rightarrow N e)$, radiative decays $(N \rightarrow \nu \gamma)$, and Dalitz-like meson decays $(\pi^0 \rightarrow \gamma \nu N)$. It also gives rise to displaced photon vertices, delayed photon signatures, and modifications of elastic recoil spectra. Recent theoretical and phenomenological studies~\cite{Magill:2018jla,Plestid:2020vqf,Brdar:2020quo} have emphasized several distinctive features: (i) the NDP produces cross sections enhanced at low recoil energies, scaling approximately as $1/E$, which gives experiments with low thresholds a significant advantage; (ii) the operator mediates radiative decays, so displaced or delayed photon signals arise naturally and serve as smoking-gun signatures; and (iii) in gauge-invariant completions the operator is embedded in higher-dimensional structures, ensuring theoretical consistency and linking dipole portal phenomenology with broader classes of models. These phenomena can be tested across an exceptionally wide range of environments, including accelerator-based neutrino experiments, colliders, reactor and solar neutrino facilities, dark matter detectors, astrophysical systems such as supernovae, and cosmology through Big Bang Nucleosynthesis (BBN) and the Cosmic Microwave Background (CMB). The fact that the same operator connects such disparate frontiers makes the NDP a powerful unifying framework for new physics searches. NDP scenarios have even been invoked to explain anomalies such as the MiniBooNE low-energy excess~\cite{MiniBooNE:2008yuf}, although such interpretations are now tightly constrained by complementary data.

Experimentally, a diverse program of searches constrains the NDP interactions. Laboratory experiments have already reached impressive sensitivities: CE$\nu$NS measurements from COHERENT~\cite{Akimov:2017ade}, electron recoil experiments such as Borexino~\cite{Agostini:2017ixy} and TEXONO~\cite{Deniz:2009mu}, and dark matter detectors like XENONnT~\cite{XENON:2024ijk} and PandaX~\cite{PandaX:2024muv} probe down to $\mu_{\nu N} \sim 10^{-10}\mu_B$. Astrophysical and cosmological constraints are even stronger: arguments from SN1987A cooling~\cite{Raffelt:1999tx}, modifications to BBN~\cite{Fields:2019pfx}, and CMB determinations of $N_\text{eff}$~\cite{Planck:2018vyg} reach $\mu_{\nu N}\lesssim 10^{-11}\mu_B$. These bounds are highly robust, being largely free of detector systematics, though they are subject to astrophysical uncertainties. Laboratory, astrophysical, and cosmological probes already exclude wide regions of parameter space, while leaving open motivated windows that can be tested in the near future.

\medskip
In this review, we present a systematic exploration of the NDP. Our aim is to integrate its theoretical foundations with its phenomenological implications, providing a coherent picture across energy, intensity, astrophysical, and cosmological frontiers. The structure of the paper is as follows.  
In Sec.~\ref{sec:framework}, we introduce the NDP operator, discuss its Lorentz and gauge structure, review possible ultraviolet completions, and summarize its main phenomenological consequences.
In Sec.~\ref{sec:production}, we examine sterile neutrino production through the dipole portal, presenting analytic expressions for cross sections and decay widths and highlighting how the operator modifies neutrino processes across up-scattering, dipole-induced meson decays, and collider channels.
In Sec.~\ref{sec:exp-sign}, we classify and review experimental search strategies, including fixed-target facilities, colliders, and recoil-based low-threshold detectors.  
Section~\ref{sec:strategy} presents a unified overview of present bounds and projected sensitivities on the NDP from laboratory, astrophysical, and cosmological probes. 
Finally, in Sec.~\ref{sec:conlcusion}, we summarize the state of the field, highlight the complementarity of existing probes, and outline promising directions for future searches. Some useful formulae are provided in the Appendix.
\section{Theoretical Framework} \label{sec:framework}
At low energies, the NDP is described by the dimension-five operator~\cite{Aparici:2009fh,Brdar:2020quo,Magill:2018jla}
\begin{equation}
\mathcal{L}_\text{NDP} \;=\; \frac{\mu_{\nu N}}{2}\,\bar{\nu}\sigma^{\mu\nu}N\,F_{\mu\nu} \;+\; \text{h.c.},
\label{eq:ndp_operator}
\end{equation}
where $\nu$ denotes an active neutrino, $N$ is a sterile gauge singlet, and $F_{\mu\nu} = \partial_\mu A_\nu - \partial_\nu A_\mu$ is the EM field strength tensor. 
The coefficient $\mu_{\nu N}$ has mass dimension $-1$ and is usually expressed in units of the Bohr magneton, $\mu_B = e/(2m_e)$, in order to facilitate comparison with neutrino magnetic moment searches. 
The bilinear $\bar{\nu}\sigma^{\mu\nu}N$ flips chirality, linking left-handed SM neutrinos to right-handed sterile states. 

Neutrino magnetic moments were first discussed in the late 1970s and early 1980s~\cite{Marciano:1977wx,Pal:1981rm,Shrock:1982sc}, originally as possible explanations for solar neutrino anomalies~\cite{Cisneros:1970nq}. 
In the SM extended with massive neutrinos, such moments arise radiatively but are extremely suppressed~\cite{Fujikawa:1980yx,Giunti:2015gga,ParticleDataGroup:2024cfk},
\begin{equation}
\mu^{\rm SM}_\nu \simeq 3 \times 10^{-19} 
\left(\frac{m_\nu}{1~{\rm eV}}\right)\mu_B ,
\end{equation}
many orders of magnitude below experimental reach. 
This motivates studying extensions such as the NDP, where transition dipoles $\mu_{\nu N}$ can be much larger, naturally arising in ultraviolet (UV) completions with new charged fermions or scalars in loops~\cite{Aparici:2009fh,Babu:2020ivd}. 
Importantly, in the limit $m_\nu \to 0$ an enhanced chiral symmetry is restored, ensuring that large transition dipoles do not destabilize the smallness of neutrino masses, in accordance with ’t~Hooft naturalness~\cite{Babu:2020ivd}.

In the SM effective field theory (SMEFT) extended with sterile fermions, the NDP arises from the higher-dimensional operator~\cite{Aparici:2009fh,Brdar:2020quo}
\begin{equation}
\frac{c}{\Lambda^2}\,(\bar L\sigma^{\mu\nu}N)\,\tilde H\,B_{\mu\nu},
\end{equation}
where $L$ is the lepton doublet, $\tilde H$ is the conjugate Higgs, $B_{\mu\nu}$ is the hypercharge field strength, and $\Lambda$ is the heavy scale. 
After electroweak symmetry breaking, this reduces to Eq.~\eqref{eq:ndp_operator} with $\mu_{\nu N}\simeq c\,v/\Lambda^2$. 
The SMEFT embedding clarifies the gauge-invariant origin of the operator and its behavior under lepton number. 
For Dirac sterile fermions with $L(N)=1$, the dipole interaction conserves lepton number, $\Delta L = 0$. 
For Majorana sterile states, only transition dipoles are allowed, while diagonal dipoles vanish identically~\cite{Shrock:1982sc,Giunti:2015gga}.

In this review, we restrict attention to transition dipoles between active and sterile states of the form $\bar\nu\sigma_{\mu\nu}N F^{\mu\nu}$. 
We do not consider sterile–sterile dipole operators ($\bar N\sigma_{\mu\nu}N F^{\mu\nu}$). 
For Majorana $N$, such diagonal dipoles are forbidden, while for Dirac $N$ they couple only sterile states to photons and do not induce active–sterile transitions. 
Although such terms could in principle induce small effective dipoles for active neutrinos in the presence of active–sterile mass mixing~\cite{Chun:2024mus,Bertuzzo:2024eds}, we do not include this possibility here, since our focus is on the minimal dipole portal without mixing effects.

At the loop level, transition dipoles can be generated with a parametric size
\begin{equation}
\mu_{\nu N} \;\sim\; \frac{e}{16\pi^2}\,\frac{m}{M^2},
\end{equation}
where $m$ denotes a chirality-flip mass insertion and $M$ the heavy mass of the loop particle. 
Explicit UV completions include models with vector-like leptons or leptoquarks~\cite{Aparici:2009fh,Babu:2020ivd,Beltran:2024twr}, where sizable $\mu_{\nu N}$ can be obtained while neutrino masses remain small. 
Such frameworks demonstrate that large dipoles can be realized consistently without destabilizing neutrino masses. 
In these cases the dipole portal may even dominate over mixing-mediated interactions, which are suppressed by small Dirac couplings or helicity suppression~\cite{Babu:2020ivd,Magill:2018jla}.

\medskip

The NDP also opens a variety of new processes of phenomenological interest:
\begin{itemize}
    \item \emph{Radiative decay:} $N \to \nu \gamma$, with width $\Gamma \propto |\mu_{\nu N}|^2 M_N^3$ where $M_N$ is a mass of HNL, often dominating over weak or mixing-induced decays for small active-sterile mixing~\cite{Pal:1981rm,Shrock:1982sc}.
     \item \emph{Neutrino up-scattering:} $\nu + {\cal A} \to N + {\cal A}$, where ${\cal A}$ can be a nucleus, proton, or electron. Cross sections receive coherent $Z^2$ enhancement for nuclear targets, making this process a powerful source of HNLs in beam-dump, reactor, and fixed-target experiments~\cite{Magill:2018jla,Plestid:2020vqf,Ismail:2021dyp}
    \item \emph{Collider production:} $e^+ e^- \to \nu N (\gamma)$ or $\gamma q \to N q$, with signatures of displaced or prompt photons plus missing energy~\cite{Ali:1974qr,Zhang:2022spf}.
    \item \emph{Recoil spectra:} Additional dipole channels modify
     $\nu$-$e$ elastic scattering (NEES) and coherent elastic neutrino-nucleus scattering (CE$\nu$NS).These do not produce $N$ directly but distort recoil distributions relative to the SM, leading to characteristic $1/E_T$ enhancements at low recoil energies~\cite{Giunti:2015gga,Brdar:2020quo,Vogel:1989iv,Miranda:2019skf}.
   \item \emph{Astrophysical and cosmological probes:} Enhanced plasmon decay rates, stellar cooling, supernova neutrino transport, and BBN constraints provide complementary sensitivity to the dipole moments~\cite{Raffelt:1996wa,Ayala:1998qz,Brdar:2020quo}.
\end{itemize}

The resulting signatures highlight the potential of the NDP to reveal new physics well beyond the SM through complementary experimental and observational probes.

\medskip


%
%
\section{Production Mechanisms}
\label{sec:production}
\subsection{Neutrino Up-scattering via the Dipole Portal}
\label{sec:dipole-upscattering}
Among the key production mechanisms for HNLs in the context of the NDP, \textit{neutrino up-scattering} stands out as both distinctive and experimentally accessible~\cite{Magill:2018jla,Ismail:2021dyp}. 
In this process, an incoming SM neutrino $\nu$ converts into a heavier state $N$ through the dipole operator with coupling $\mu_{\nu N}$, mediated by the exchange of a virtual photon. 
The generic form of the reaction is
\begin{equation}
\nu(p) + {\cal A}(k) \to N(p') + {\cal A}(k'),
\end{equation}
where ${\cal A}$ denotes the target, which may be a nucleus, proton, or electron. The momentum transfer is characterized by $t = (p - p')^2$, and the dynamics are governed by the effective dipole interaction.

\medskip

The neutrino up-scattering is especially relevant at the intensity frontier, such as beam-dump facilities and near-detector experiments, where intense neutrino fluxes interact with dense targets. 
The production probability increases with higher incoming neutrino energy $E_\nu$ and with larger HNL mass $M_N$ (as long as kinematics permit), since the available phase space expands. 
This makes the neutrino up-scattering a powerful probe of the NDP across a broad range of energies and masses.

The scattering dynamics depend on the target composition and the momentum transfer, which broadly divides the interaction into two regimes:
\begin{itemize}
\item \textbf{Coherent scattering:} At low momentum transfer $|t|$, neutrinos scatter coherently off the entire nucleus and, at very low energies, even off the atomic electron cloud. 
For a nucleus with atomic number $Z$ (protons) and neutron number $N$, the cross section acquires a coherence enhancement proportional to the square of the net charge distribution: approximately $N^2$ for neutral–current weak interactions, or $Z^2$ for purely EM interactions.
This enhancement is reduced at higher $|t|$ by nuclear form factors~\cite{Freedman:1973yd,Scholberg:2005qs,Bednyakov:2018mjd}. 
Coherent scattering dominates at low neutrino energies (tens of MeV) and small scattering angles.
\item \textbf{Incoherent scattering:} At higher $E_\nu$ or larger $|t|$, the neutrino resolves individual nucleons or electrons. 
In this regime, the cross section scales only linearly with the number of scattering centers ($\propto Z$ for protons or electrons, $\propto A (\text{mass number})$ for all nucleons), lacking the collective $A^2$ or $N^2$ coherence enhancement.  
Incoherent scattering becomes dominant once nucleon recoil or nuclear breakup is kinematically accessible~\cite{Bednyakov:2018mjd,Bednyakov:2021bty}.
\end{itemize}

\medskip

The target type determines which regime dominates in practice:
\begin{itemize}
    \item \textbf{Nuclear targets} support both coherent and incoherent interactions.
    At low momentum transfer, coherence leads to an enhancement scaling as $\sim N^2$ for neutral–current weak interactions or $\sim Z^2$ for EM dipole interactions, while incoherent processes dominate at higher energies or large $|t|$~\cite{Bednyakov:2018mjd}.
    \item \textbf{Free protons}, as in hydrogen-rich materials, provide clean elastic up-scattering channels over a wide energy range. Inelastic contributions from bound nucleons, including nuclear excitations and breakup, become relevant in dense nuclear environments~\cite{MiniBooNE:2010xqw,Duyang:2018xcc}.
    \item \textbf{Electrons} offer purely elastic, point-like scattering without form-factor suppression. Although the cross section is smaller due to the electron’s low mass and absence of nuclear enhancement, this channel remains important in low threshold detectors with large electron densities~\cite{Radel:1993sw,Tomalak:2019ibg}.
\end{itemize}

\medskip

Overall, the neutrino up-scattering via the NDP is accessible over a wide range of energies and target materials. 
Its characteristic experimental signatures such as forward-peaked kinematics, displaced decays, and mono-photon final states make it  a particularly promising channel for probing the NDP at both current and next-generation facilities.

\begin{figure}[!ht]
\centering
\begin{center}
\includegraphics[width=
5cm]{
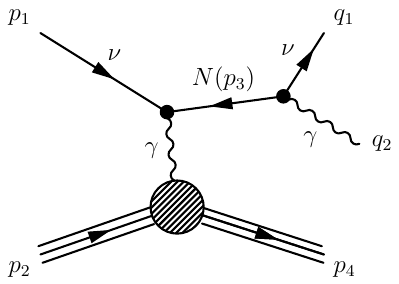}
\end{center}
\caption{Tree level neutrino scattering process with a final state photon, arising from dipole portal to HNL~\cite{Magill:2018jla}. 
}
\label{fig:signalfeynmandiagram}
\end{figure}
Dipole-induced differential cross sections for HNL production play a central role in describing neutrino up-scattering processes.
They depend on both the recoil energy $E_r$ of the final state target and the energy $E_\nu$ of the incoming neutrino. 
In general, the differential cross section is proportional to the contraction of a leptonic tensor $L_{\mu\nu}$ with a hadronic tensor $W^{\mu\nu}$. 
In terms of the right-handed projection operator, the leptonic tensor takes the form~\cite{Magill:2018jla}
\begin{align}
L_{\mu\nu}=4\,{\rm Tr}\!\left[\slashed{p}_1 P_R\sigma_{\nu\alpha}q^{\alpha}(\slashed{p}_3+m_N)\sigma_{\mu\beta} q^{\beta}\right],
\end{align}
while the hadronic current is expressed as
\begin{align}
\langle {\cal A}|\Gamma^{\mu}|{\cal A}^{\prime}\rangle
=F_1 \gamma^{\mu}
+F_2\frac{i}{2 M_H}\sigma^{\mu\delta}q_{\delta},
\label{eq:hc}
\end{align}
where $F_1$ and $F_2$ are form factors and $M_H$ is the hadronic mass. 

\medskip

In the coherent regime, the neutrino scatters off the entire nucleus as a single object. 
The cross section then scales approximately as $Z^2$ (or $A^2$), and since $M_H = A M_{\rm nucleon}$ with $|t|=|q^2|=Q^2$ small, only the $F_1$ term is retained in Eq.~\eqref{eq:hc}. 
In contrast, inelastic (incoherent) scattering occurs when the neutrino interacts with individual nucleons rather than the nucleus as a whole. 
In this case, $|t|$ is of moderate size, we set $M_H = M_{\rm nucleon}$, and both $F_1$ and $F_2$ contribute. 
The form factors differ for protons and neutrons, with numerical values provided in Refs.~\cite{Perdrisat:2006hj,Beck:2001dz,Magill:2018jla}. 
Unlike the coherent case, the inelastic rate scales only linearly with $Z$ or $A$, and the relevant kinematics avoid the forward scattering ($t\!\to\!0$) enhancement characteristic of coherent scattering.

\medskip

The explicit cross sections differ for electron and nuclear targets~\cite{Brdar:2020quo}.  
For electron scattering $(\nu e \to N e)$ one finds
{\small
\begin{align}
  \frac{d\sigma_{\mu}}{dE_r}
    &= \alpha \mu_{\nu N}^2 \bigg[
         \frac{1}{E_r} - \frac{1}{E_\nu}
       + M_N^2 \frac{E_r - 2 E_\nu - m_e}{4 E_\nu^2 E_r m_e}
       + M_N^4 \frac{E_r - m_e}{8 E_\nu^2 E_r^2 m_e^2}
       \bigg],
      \label{eq:sigma-mm-e}
\end{align}
}
while for nuclear targets $X_Z^A$ $(\nu ~X_Z^A \to N~ X_Z^A)$~\cite{Brdar:2020quo},
{\small
\begin{align}
  \frac{d\sigma_{\mu}}{dE_r}
   &= \alpha \mu_{\nu N}^2 Z^2 F_1^2(E_r) \bigg[
         \frac{1}{E_r} - \frac{1}{E_\nu}
       + M_N^2 \frac{E_r - 2 E_\nu - m_X}{4 E_\nu^2 E_r m_X}
       + M_N^4 \frac{E_r - m_X}{8 E_\nu^2 E_r^2 m_X^2}
       \bigg] \nonumber\\
&+\alpha \mu_{\nu N}^2 \mu_X^2 F_2^2(E_r) \bigg[
         \frac{2m_X}{E_r^2}\!\left((2E_{\nu}-E_r)^2-2E_r m_X \right)
       + M_N^2 \frac{E_r - 4 E_\nu }{E_\nu^2}
       + M_N^4 \frac{1}{E_\nu^2 E_r}
       \bigg],
  \label{eq:sigma-mm-N}
\end{align}
}
where $M_N$ is the HNL mass, $\alpha$ the EM fine-structure constant, $\mu_X$ the nuclear magnetic moment, $\mu_{\nu N}$ the transition dipole moment, and $m_X$ the target mass. 
At higher energies, nuclear substructure is partially resolved and coherence is lost, in which case both $F_1$ and $F_2$ are required. 
Magnetic moment contributions are typically negligible compared to the coherent $Z^2$-enhanced term at low energies. 
For the nuclear charge form factor $F_1(E_r)$, we adopt the Helm parameterization~\cite{Helm:1956zz}, which is practically equivalent to the symmetrized Fermi~\cite{Piekarewicz:2016vbn} and Klein–Nystrand~\cite{Klein:1999qj} forms. 
\footnote{While the Helm form is widely used for its simplicity, more accurate descriptions—such as Woods–Saxon~\cite{Woods:1954zz} or Fourier–Bessel expansions—are recommended for heavy nuclei and experiments sensitive to detailed recoil spectra~\cite{Duda:2006uk}.} 

\medskip

The Helm form factor is given by~\cite{Helm:1956zz}
\begin{align}
F_{\rm Helm}(|\vec{q}|^2) 
&= 3\,\frac{j_1(|\vec{q}|\,R_0)}{|\vec{q}|\,R_0} \;
e^{-|\vec{q}|^2 s^2/2},
\label{eq:Helm-FF}
\end{align}
where $j_1(x) = (\sin x - x \cos x)/x^2$ is the spherical Bessel function of order one, 
$|\vec{q}| = \sqrt{2 m_{X} E_r}$ is the magnitude of the three-momentum transfer for a recoil of energy $E_r$, and $s$ is the nuclear skin thickness (typically $s=1.0~\mathrm{fm}$).  
The diffraction radius $R_0$ relates to the effective nuclear radius $R = 1.2\,A^{1/3}~\mathrm{fm}$ by $R_0 = \sqrt{R^2 - 5 s^2}$.

\medskip

The kinematics further constrain the recoil spectrum. 
The minimum neutrino energy required to produce a recoil of energy $E_r$ is
\begin{align}
  E_\nu^\text{min}(E_r) 
  = \tfrac{1}{2} \Big[ E_r + \sqrt{E_r^2 + 2 m_X E_r} \Big]
    \left(1 + \frac{M_N^2}{2 E_r m_X} \right),
  \label{eq:E-nu-min}
\end{align}
while the recoil spectrum vanishes below
\begin{align}
  E_r^\text{min} = \frac{M_N^2}{2( m_X + M_N)}.
  \label{eq:Er-min}
\end{align}
For fixed $E_\nu$, the recoil spectrum peaks at
\begin{align}
  E_r^\text{peak}(E_\nu)
    = \frac{2 m_X M_N^4}{8 E_\nu^2 m_X^2 - 2 m_X M_N^2 (2 E_\nu + m_X) + M_N^4},
  \label{eq:Er-peak}
\end{align}
and the maximum recoil energy is
\begin{eqnarray}
  E_r^\text{max}(E_\nu)
    &=& \frac{1}{2 E_\nu + m_X}
      \bigg[
        E_\nu^2 - \tfrac{1}{2} M_N^2
      + \frac{E_\nu}{2 m_X} \Big(
            \sqrt{M_N^4 - 4 M_N^2 m_X (E_\nu + m_X) + 4 E_\nu^2 m_X^2} \nonumber \\
&& - M_N^2 \Big)
      \bigg].
  \label{eq:Er-max}
\end{eqnarray}

\medskip

From these relations, one can see that producing a HNL via up-scattering off electrons requires a higher incoming neutrino energy than off nuclei of the same HNL mass. 
This is because the light electron absorbs a larger fraction of the recoil energy, reducing the phase space available for HNL production~\cite{Li:2022bqr}.

When the neutrino resolves individual nucleons rather than the entire nucleus, coherence is lost and the scattering becomes inelastic~\cite{Bednyakov:2018mjd,Bednyakov:2021bty}. In this regime, the momentum transfer $|t|$ is of moderate size, and the cross section scales linearly with the nucleon number ($\propto A$). The dynamics are governed by the nucleon form factors $F_1$ and $F_2$, with the magnetic contribution in $F_2$ usually subdominant. Inelastic processes may excite discrete nuclear levels or lead to breakup into continuum states, though the rate is suppressed whenever the excitation energy $\Delta E \gtrsim E_r$. At sufficiently high neutrino energies, incoherent scattering on individual nucleons dominates. Explicit cross-section expressions are provided in Refs.~\cite{Vogel:1989iv,Bednyakov:2018mjd}, while more detailed treatments of nuclear response functions and structure effects can be found in Refs.~\cite{Formaggio:2012cpf,Giunti:2015gga}.


The neutrino up-scattering is a leading production mechanism for HNL across wide regions of parameter space and has been probed using data from LSND~\cite{LSND:1996ubh}, MiniBooNE~\cite{MiniBooNE:2018esg}, and the NOMAD single-photon search~\cite{NOMAD:2011gyy}, which have been recast to constrain dipole-mediated production and $N \to\nu\gamma$ decays. Future high intensity facilities, including SHiP~\cite{SHiP:2015vad} and the Short-Baseline Neutrino program (SBN)~\cite{MicroBooNE:2015bmn}, are projected to achieve strong sensitivity~\cite{Magill:2018jla}.
Complementary bounds on the dipole coupling $\mu_{\nu N}$ arise from precision NEES, with results from GEMMA~\cite{Beda:2012zz,Beda:2013mta}, TEXONO~\cite{TEXONO:2006xds,TEXONO:2009knm}, Borexino~\cite{Agostini:2017ixy}, and XENON1T~\cite{XENON:2020rca}. Additional constraints come from CE$\nu$NS, observed by COHERENT~\cite{COHERENT:2020iec} and under active study at reactor-based experiments such as CONNIE~\cite{CONNIE:2021ggh} and CONUS~\cite{CONUS:2020skt}.

%
\subsection{Meson-Induced Production}
\label{sec:meson-decays}
\begin{figure}[t]
    \centering
    \includegraphics[width=\textwidth]{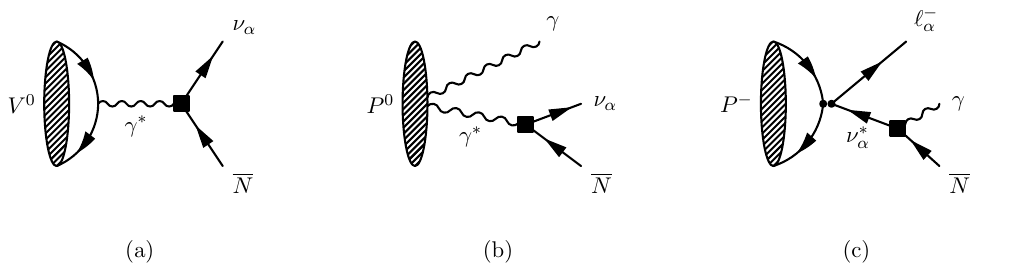}
    \caption{Representative meson decay topologies induced by the NDP, taken from Ref.~\cite{Barducci:2023hzo} : (a) Two-body vector meson decay, (b) Radiative Dalitz-like neutral meson decay, and (c) Weak three-body charged meson decay.}
    \label{fig:decays}
\end{figure}

In addition to neutrino up-scattering, HNLs can also be produced in meson decays via the NDP, providing unique and complementary channels beyond those associated with active–sterile mass mixing. 
These processes are particularly relevant in high intensity environments with abundant meson production, where both prompt and long-lived mesons may contribute depending on the experimental configuration.
Dipole-induced meson decays can be classified into three main categories~\cite{Barducci:2023hzo}:
\begin{itemize}
    \item Two-body decays of vector mesons,
    \item Radiative Dalitz-like decays of neutral pseudoscalar mesons~\cite{Kroll:1955zu,Mikaelian:1972yg,Landsberg:1985gaz},
    \item Weak three-body decays of charged pseudoscalar mesons via off-shell neutrinos.
\end{itemize}
Each class corresponds to a distinct topology mediated by the dipole operator, as illustrated in Fig.~\ref{fig:decays}, and exhibits characteristic kinematics and flavor dependence.

The relative importance of these modes depends on the experimental setup, in particular on how mesons are treated after production. 
In decay-pipe experiments (e.g.\ MiniBooNE~\cite{MiniBooNE:2018esg}, SBN~\cite{MicroBooNE:2015bmn,Dutta:2025npn}), long-lived charged mesons such as $\pi^\pm$ and $K^\pm$ dominate, since their decays in flight (DIFs) produce energetic HNLs that can reach downstream detectors. 
By contrast, short-lived neutral mesons ($\pi^0$, $\eta$) decay promptly at the target. Although they can contribute through radiative dipole channels (e.g.\ $\pi^0 \to \gamma N \nu$), their limited boost and strongly forward-peaked final states typically make them subleading. 
In beam-dump experiments~\cite{Kanemura:2015cxa,Giffin:2022rei}, absorption of long-lived charged mesons suppresses this contribution, shifting sensitivity toward prompt neutral meson decays. 
At colliders, heavy mesons such as $D$ and $B$ and light vector mesons ($\rho^0$, $\omega$, $\phi$) are accessible, where rare dipole-induced decays in these channels can be reconstructed with high precision~\cite{BESIII:2011ysp,Belle-II:2010dht,Tiwary:2024vit}. 
The specific realizations of these three experimental configurations, decay-pipe, beam-dump, and collider setups, will be discussed in detail in the following sections.
We now turn to specific meson decay modes that can produce HNLs via the dipole portal interaction.
\subsubsection{Two-Body Decay of Vector Mesons}
Flavorless vector mesons ($V^{0}$) like $\rho^0$, $\omega$, $\phi$, $J/\psi$, and $\Upsilon(1S)$ can decay directly into a neutrino and an HNL via the dipole operator as shown in Fig. \ref{fig:decays}(a):
\begin{equation}
V^0 \to \nu_\alpha N.
\end{equation}
This provides a clean and kinematically favorable channel to probe the NDP~\cite{Barducci:2023hzo}.
For a vector meson, one has
\begin{equation}
 \langle 0 | \overline{q_i} \gamma^\mu q_j | V^0(p,\epsilon) \rangle = i f_V m_V \epsilon^\mu\,, 
\end{equation}
where $m_V$ and $f_V$ are the mass and the decay constant of $V^0$, respectively, and $\epsilon^\mu$ is its polarization vector. 
The decay width of interest reads~\cite{Barducci:2023hzo}
\begin{equation}
 \Gamma\left(V^0 \to \nu_\alpha N \right) = \frac{f_V^2 m_V}{12\pi} 
 e^2 Q_q^2 \left|\mu_{\nu N}\right|^2  
 \left(1-\frac{M_N^2}{m_V^2}\right) \left(1+\frac{M_N^2}{m_V^2}-2\frac{M_N^4}{m_V^4}\right)\,,
 \label{eq:GammaV0nuNDir}
\end{equation}
where  
$e$ is the electron charge, and $Q_q$ is the charge (in units of $e$) of quark $q$ composing $V^0$. 
For light vector mesons such as $\rho^0$, $\omega$, and $\phi$, the branching ratios are extremely small, $\mathrm{BR} \lesssim 10^{-14}$ for $|\mu_{\nu N}| = 10^{-6}~\mathrm{GeV}^{-1}$, owing to the combination of modest decay constants, EM coupling suppression, and large total widths~\cite{Barducci:2023hzo}.
As a result, the expected event yields in neutrino beam-dump or fixed-target experiments are negligible compared to dominant production channels such as neutrino up-scattering and radiative (Dalitz-like) pseudoscalar meson decays, which benefit from coherent $Z^2$ enhancement or large meson yields, respectively.
This channel becomes potentially relevant only in environments with abundant heavy quarkonium production, such as forward LHC experiments (FASER/FASER2~\cite{FASER:2022hcn}, FACET~\cite{Cerci:2021nlb}), where narrow widths of states like $J/\psi$ and $\Upsilon$ reduce the branching ratio suppression, or in scenarios where 
the HNL mass kinematically forbids lighter meson channels.

\subsubsection{Radiative Dalitz-like Decays of Neutral Mesons}
 Neutral pseudoscalar mesons $P^0$ (e.g.\ $\pi^0$, $\eta$) decay into two photons via the chiral anomaly.
If one photon is off-shell, it can convert through the dipole operator into $\nu N$, leading to the three-body process
$P^0 \!\to\! \gamma\,\nu N$ as shown in Fig.~\ref{fig:decays}(b)~\cite{Magill:2018jla,Barducci:2023hzo}.
This channel is directly induced by the NDP and is closely analogous to the standard Dalitz decay $P^0\!\to\!\gamma\,\ell^+\ell^-$~\cite{Kroll:1955zu,Mikaelian:1972yg,Landsberg:1985gaz}.
A convenient representation of the amplitude is~\cite{Barducci:2023hzo}
\begin{equation}
 i\mathcal{M} \;=\;
 \frac{e^2}{2\pi^2}\,\frac{\mu_{\nu N}}{f_P}\;
 \epsilon_\mu^\ast\,\epsilon^{\mu\nu\alpha\beta}\;
 \frac{g_{\nu\sigma}}{q^2}\;
 p_{2\alpha}\,q_\beta\,q_\rho\;
 \big[\overline{u_\nu}\,\sigma^{\rho\sigma} P_R v_N\big],
\end{equation}
where $f_P$ is the meson decay constant, $p_2$ is the momentum of the on-shell photon, and $q\!\equiv\!p_1\!-\!p_2$ is the momentum of the virtual photon.
Expressed in terms of the angle $\theta$ between the $N$ and the real photon in the $q$–rest frame, the squared amplitude and the double differential width read~\cite{Barducci:2023hzo}
\begin{eqnarray}
 \big|\mathcal{M}\big|^2
 = \frac{e^4}{32\pi^4}\,
    \frac{|\mu_{\nu N}|^2}{f_P^2}\,
    \frac{(q^2-m_P^2)^2\,(q^2-M_N^2)}{q^4}\,
    \Big[q^2+3M_N^2-\big(q^2-M_N^2\big)\cos 2\theta\Big], 
     \end{eqnarray}
 \begin{eqnarray}
 \frac{d^2\Gamma(P^0\to\gamma\,\nu N)}{dq^2\,d\cos\theta}
 = \frac{1}{512\pi^3 m_P}\,
    \Big(1-\frac{q^2}{m_P^2}\Big)\!
    \Big(1-\frac{M_N^2}{q^2}\Big)\;
    \big|\mathcal{M}\big|^2.
\end{eqnarray}
After integrating $q^2\!\in\![M_N^2,m_P^2]$ and $\cos\theta\!\in\![-1,1]$, one obtains the expected scaling
\(
\Gamma(P^0\!\to\!\gamma\nu N)\!\sim\!
 \alpha^2\,\mu_{\nu N}^2\,m_P^3\,
 I(0,M_N^2/m_P^2),
\)
with a three-body phase space function $I(0,M_N^2/m_P^2)$ given by Eq.(\ref{eq:loop-func}).   
Because the final-state photon is massless, the spectrum is forward-peaked and the mode remains sensitive up to the kinematic limit $M_N\simeq m_P$.

\medskip

These Dalitz-like decays are most effectively probed in environments with abundant $P^0$ production and efficient photon reconstruction.
Fixed-target kaon experiments (e.g.\ NA62) provide copious $\pi^0$ from $K^+\!\to\!\pi^+\pi^0$ decays~\cite{NA62:2017rwk,NA62:2023olg}, while collider detectors (Belle~II, LHCb) can exploit high statistics samples of $\eta$, $\eta'$, and $P^0$ from heavy flavour and light hadron production, with excellent photon and vertexing performance~\cite{BelleII:TDR,LHCb:VELO,AbellanBeteta:2020amj}.
\subsubsection{Weak Three-Body Decays via Off-Shell Neutrino}
Charged pseudoscalar mesons ($P^\pm$), such as $\pi^\pm$, $K^\pm$, $D^\pm$, $D_s^\pm$, and $B^\pm$, can decay into a charged lepton, photon, and HNL via an off–shell virtual neutrino, as shown in Fig.~\ref{fig:decays}(c)~\cite{Magill:2018jla,Barducci:2023hzo}:
\begin{equation}
P^\pm \;\to\; \ell_\alpha^\pm \,\big(\nu_\alpha^* \to \gamma N\big).
\end{equation}
This process proceeds through a charged–current weak interaction followed by a dipole mediated transition.  
It is flavor specific, enhanced for $\nu_\mu$ due to the high flux of muon neutrinos in accelerator experiments, and is most relevant for low mass HNLs with $M_N \lesssim m_P - m_{\ell_\alpha}$.  
Such decays are loop–suppressed in the SM and are described via effective dipole operators.  
They have been discussed in the context of pion and kaon production in LSND, MiniBooNE, and SBN~\cite{Magill:2018jla}.

The process can be viewed as a rare three–body decay of the form
\[
P^-(q) \;\to\; \ell^-(p_\ell)\,\gamma(k)\,N(p_N),
\]
induced by an off–shell active neutrino transitioning into an HNL ($N$) via the dipole operator.  
The amplitude for $P^- \to \gamma \ell^- N$ is given schematically by~\cite{Barducci:2023hzo}
\begin{equation}
 i\mathcal{M} = i\,2\sqrt{2}\,G_F\,V_{ij}\,f_P\,\mu_{\nu N}\,
 \frac{1}{q^2}\,k_{1\alpha}\,\epsilon_\beta^\ast 
 \Big[\overline{u_\ell}\,\slashed{p}_1\slashed{q}\,\sigma^{\alpha\beta} P_R v_N\Big],
 \label{eq:PgammalN}
\end{equation}
where $G_F$ is the Fermi constant, $V_{ij}$ is the relevant CKM element for $P^-\sim\bar u_i d_j$, and $f_P$ is the decay constant defined through
\[
 \langle0|\bar u_i \gamma^\mu\gamma_5 d_j|P^-(p_1)\rangle = i f_P p_1^\mu.
\]
Squaring the amplitude and expressing the scalar products in terms of the angle $\theta$ between the HNL and the charged lepton ($l_\alpha$) in the virtual neutrino rest frame, one finds~\cite{Barducci:2023hzo}
\begin{align}
 |\mathcal{M}|^2 &= 8 G_F^2 |V_{ij}|^2 f_P^2 |\mu_{\nu N}|^2 
 \frac{(q^2-M_N^2)^2}{q^4}\,\Big[q^2(m_P^2-q^2)
 +(2q^2+m_P^2)m_{\ell_\alpha}^2 - m_{\ell_\alpha}^4 \nonumber\\
 &\qquad + (q^2-m_{\ell_\alpha}^2)\,\sqrt{\lambda(q^2,m_P^2,m_{\ell_\alpha}^2)}\cos\theta\Big],
\end{align}
with $\lambda$ the Källén function given by Eq.(\ref{kallen}).  
The differential decay rate is then~\cite{Barducci:2023hzo}
\begin{equation}
 \frac{d^2\Gamma(P^- \to \gamma \ell_\alpha^- N)}{dq^2\,d\cos\theta}
 = \frac{\sqrt{\lambda(q^2,m_P^2,m_{\ell_\alpha}^2)}}{512\pi^3 m_P^3}
 \Big(1-\frac{M_N^2}{q^2}\Big)\,|\mathcal{M}|^2,
\end{equation}
with $q^2 \in [M_N^2,(m_P-m_{\ell_\alpha})^2]$ and $\cos\theta\in[-1,1]$.

For charged pions, the integrated three–body decay width can be approximated as
\begin{equation}
\Gamma(\pi^- \to \mu^- \gamma N) \;\sim\;G_F^2\,f_\pi^2\,|V_{ud}|^2\,|\mu_{\nu N}|^2\,m_\pi^5\,I(r_\mu,r_N),
\end{equation}
where $r_\mu = m_\mu^2/m_\pi^2$, $r_N = M_N^2/m_\pi^2$.
This scaling highlights that the dipole operator lifts the helicity suppression present in SM processes such as $\pi^- \to e^- \bar{\nu}$.  
The corresponding SM two–body rate is
\[
\Gamma(\pi^- \to \mu^- \bar{\nu}) = \frac{G_F^2 |V_{ud}|^2 f_\pi^2 m_\pi m_\mu^2}{8\pi}
 \Big(1-\frac{m_\mu^2}{m_\pi^2}\Big)^2,
\]
and the approximate branching ratio can be written as
\begin{equation}
{\rm BR}(\pi^- \to \mu^- \gamma N)\;\sim\;
 \frac{|\mu_{\nu N}|^2\,m_\pi^4}{m_\mu^2}\;
 \frac{I(r_{\mu},r_N)}{(1-r_N)^2}.
\end{equation}
Thus, the branching ratio scales quadratically with the dipole strength and is suppressed as $M_N\!\to\!m_\pi-m_\mu$.

Charged pseudoscalar mesons can also undergo rare \emph{four-body} decays in which a virtual photon converts to a charged lepton pair via the dipole operator. 
The generic process is
\[
M^+(q) \;\to\; \mu^+(p_\mu)\;\ell^+(p_+)\;\ell^-(p_-)\;N(p_N),
\]
mediated by an off-shell neutrino $\nu^*$ that transitions into a HNL through the dipole interaction and radiates a virtual photon $\gamma^*\!\to\ell^+\ell^-$~\cite{Barducci:2023hzo}. 
This channel is universal across neutrino flavors and is relevant for HNL production in high-intensity facilities such as LSND, MiniBooNE, SBN, and SHiP.
At leading order, the amplitude takes the schematic form
\begin{eqnarray}
\mathcal{M}
&\sim\frac{G_F}{\sqrt{2}}\,V_{qq'}\,f_M\,\mu_{\nu N}\;
\Big[\overline{u}_\mu(p_\mu)\,\gamma^\alpha(1-\gamma^5)\,
\frac{\slashed{p}_\nu}{p_\nu^2}\,
\sigma_{\rho\lambda}\,q^{\prime\,\lambda}P_R\,
u_N(p_N)\Big]
\\
&\hspace{2.3cm}\times\;
\frac{-g^{\rho\beta}}{q^{\prime\,2}}\;
\Big[\overline{u}_-(p_-)\,\gamma_\beta\,v_+(p_+)\Big],
\label{eq:Mto_mullN_amp}
\end{eqnarray}
where $p_\nu=q-p_\mu$ is the off-shell neutrino momentum, $q'=p_+ + p_-$ the virtual photon momentum, and $\mu_{\nu N}$ is absorbed into the dipole vertex. 
The term $\bar u_- \gamma_\mu v_+$ represents the $\ell^+\ell^-$ current. 
The complete four-body differential decay rate is given in~\ref{sec:4-body}.

Introducing the invariant mass of the lepton pair, $m_{\ell\ell}^2=q'^2=(p_++p_-)^2$, the leading dependence of the spectrum can be written as
\begin{equation}
\frac{d\Gamma}{dm_{\ell\ell}^2} \;\propto\; 
\frac{G^2_F f^2_M|\mu_{\nu N}|^2 m^3_M}{m_{\ell\ell}^2}\,
\lambda^{1/2}\!\left(m_M^2,\,m_\mu^2,\,M_N^2+m_{\ell\ell}^2\right),
\end{equation}
up to spinor-trace and angular factors. 
Here $\lambda$ is the Källén function. 
The rate is enhanced at small $m_{\ell\ell}$ due to the $1/q'^2$ photon propagator, while the phase space closes at the endpoint 
$m_M < m_\mu+M_N+2m_\ell$. 
The $m_{\ell\ell}$ spectrum is therefore sharply peaked near threshold, $m_{\ell\ell}\simeq 2m_\ell$, mimicking an ordinary Dalitz decay~\cite{Kroll:1955zu,Mikaelian:1972yg,Landsberg:1985gaz}.
The differential full four-body decay rate is defined in~\ref{sec:4-body}.
A rough parametric estimate of the branching ratio (neglecting angular integrals) can be obtained by treating the four-body mode as the three-body radiative decay
multiplied by the probability for $\gamma^\ast\!\to\ell^+\ell^-$ conversion. 
Following the analysis of Dalitz decays~\cite{Kroll:1955zu,Mikaelian:1972yg,Landsberg:1985gaz}, one finds
\begin{equation}
\mathrm{BR}(M\to\mu\,\ell^+\ell^- N)
\;\sim\; 
\frac{\alpha} {3\pi}\,
\mathrm{BR}(M\to\mu\,\gamma N)\,
\log\!\left(\frac{m_M^2}{m_\ell^2}\right),
\end{equation}
where the logarithm reflects the collinear enhancement in $\gamma^\ast\!\to\ell^+\ell^-$,
regulated by the lepton mass $m_\ell$. 
In contrast to the three-body radiative mode, which yields a broad photon energy spectrum, 
the four-body channel exhibits a sharp peak in the dilepton invariant-mass distribution 
near $m_{\ell\ell}\simeq 2m_\ell$.

\subsection{Production at Colliders}
Collider experiments provide a complementary probe of the NDP, offering clean kinematics and extending sensitivity to higher HNL masses than fixed-target facilities. 
At the partonic level, the leading dipole–mediated process is
\begin{equation}
e^+e^- \;\to\; \gamma^\ast \;\to\; \nu N \;(\text{or } \bar\nu N),
\end{equation}
with a differential cross section~\cite{Zhang:2022spf} 
\begin{equation}
\frac{d\sigma}{d\cos\theta_N}
=\frac{\mu^2_{\nu N}\,\alpha \,(s-M_N^2)^2
\big[(1-\cos^2\theta_N)\,s+(1+\cos^2\theta_N)\,M_N^2\big]}{4s^3},
\end{equation}
where $\theta_N$ is the HNL scattering angle. 
The total cross section is
\begin{equation}
\sigma(e^+e^- \to \nu N)
=\frac{\mu^2_{\nu N}\,\alpha \,(s-M_N^2)^2\,(s+2M_N^2)}{3s^3}.
\end{equation}

The two-body final state $\nu N$ is experimentally invisible, being indistinguishable from the SM background $e^+e^- \to \nu\bar\nu$. 
Searches therefore focus on the radiative channel
\begin{equation}
e^+ e^- \;\to\; \nu N \gamma ,
\end{equation}
where the photon is emitted either from the initial state or directly at the dipole vertex. 
The additional photon provides a clean trigger and efficiently suppresses the otherwise overwhelming $e^+ e^- \to \nu \bar{\nu}$ background. 
Facilities such as Belle~II~\cite{Belle-II:2010dht} and BESIII~\cite{BESIII:2011ysp} are sensitive to HNLs with $m_N \sim 10$–$500$~MeV, probing both mono-photon plus missing energy $(\gamma+\slashed{E}_T)$ signatures and displaced photons from long-lived $N$ decays.

At LEP, mono-photon searches for $e^+e^- \to \gamma+\slashed{E}_T$ were carried out by ALEPH and OPAL, setting limits on anomalous neutral-current processes and thereby constraining dipole couplings up to $M_N$ near the electroweak scale~\cite{ALEPH:1993pqw,OPAL:2000puu}. 
At the LHC ($\sqrt{s}=13$–$14$~TeV), the analogous process arises via Drell–Yan production,
\begin{equation}
q\bar q \;\to\; \gamma^\ast \;\to\; \nu N ,
\end{equation}
followed by $N \to \nu\gamma$, which yields the characteristic $\gamma+\slashed{E}_T$ topology. 
ATLAS and CMS have performed mono-photon searches exploiting isolated high-$p_T$ photons accompanied by large missing energy to suppress SM backgrounds~\cite{ATLAS:2017pzz,CMS:2017qyo}. 
Subleading photon–initiated processes such as $\gamma q \to N q$ may also contribute but are suppressed, while $\gamma\gamma \to \nu N$ requires two dipole insertions and is negligible compared to the Drell–Yan channel.

In summary, collider production probes the NDP through electroweak processes
tagged by photons: Belle~II~\cite{Belle-II:2010dht} and BESIII~\cite{BESIII:2011ysp}
cover the sub-GeV mass window, LEP~\cite{ALEPH:1993pqw,OPAL:2000puu} constrained
up to tens of GeV, and the LHC~\cite{ATLAS:2017pzz,CMS:2017qyo} extends
sensitivity into the multi-GeV regime.

\section{Experimental Signatures}
\label{sec:exp-sign}
Experimental probes of the NDP rely on two qualitatively different classes of observable signatures. 
First, if the HNL is produced on shell, it will eventually decay radiatively via $N \to \nu\gamma$, yielding a photon signal that may appear promptly, with a displaced vertex, or with a measurable time delay. 
These decay-based signatures are the hallmark of beam-dump, fixed-target, and collider experiments searching for long-lived particles. 
Second, even in the absence of visible decays, dipole interactions can manifest as distortions in recoil energy spectra from elastic scattering processes such as NEES or CE$\nu$NS. 
In this case, the signal consists of an excess of low energy recoils relative to the  expectation, driven by the characteristic $1/E_r$ enhancement of the dipole contribution~\cite{Vogel:1989iv,Giunti:2015gga}. 
We then expect that these complementary strategies, through both decay signatures and recoil distortions, cover a wide range of lifetimes, energies, and detector technologies, providing a unified framework to test the NDP.
\subsection{Radiative Deacy Signatures}
The produced HNL decays radiatively via $N \to \nu \gamma$, with width~\cite{Pal:1981rm} 
\begin{equation}
\Gamma(N \to \nu \gamma) = \frac{|\mu_{\nu N}|^2 M_N^3}{4\pi}.
\end{equation}
This dominant two-body decay produces an active neutrino and a photon. 
The corresponding lab-frame decay length is~\cite{Magill:2018jla}
\begin{equation}
L_{\rm dec}=\frac{E_N}{M_N}\cdot \frac{1}{\Gamma(N\to\nu\gamma)}
=\frac{4\pi E_N}{|\mu_{\nu N}|^2 M_N^4},
\end{equation}
where $E_N$ is the HNL energy. 
For example, taking $E_N=100$~GeV, $M_N=150$~MeV, and $\mu_{\nu N}=10^{-6}$~GeV$^{-1}$ gives $L_{\rm dec}\sim 490$~m.

In high-density detector, the signal photon typically converts into a detectable $e^+e^-$ pair, while in low-density detectors the rare three-body decay $N \to \nu e^+e^-$ (via a virtual photon, branching ratio $\sim 0.6\%$~\cite{Arguelles:2021dqn}) may be easier to identify owing to reduced neutrino-induced backgrounds. 
Depending on the dipole coupling and the HNL mass, the decay can be prompt or occur with a displaced vertex inside the detector. 
The $N \to \nu \gamma$ mode can therefore manifest experimentally as:
\begin{itemize}
\item \textbf{Prompt decays:} Photons appear at the primary vertex with no resolvable displacement ($L\ll$ detector resolution). These are kinematically indistinguishable from SM backgrounds such as $\pi^0\to\gamma\gamma$, bremsstrahlung, or radiative meson decays. Experimental handles include parent-meson tagging (e.g. $K^+\to \ell^+N$), mono-photon plus missing energy searches, deviations in photon spectra, or fast timing. Experiments with clean environments and kinematic control (NA62~\cite{NA62:2017rwk}, LSND~\cite{LSND:1996ubh}, MiniBooNE~\cite{MiniBooNE:2018esg}, Belle~II~\cite{Belle-II:2010dht}) are especially sensitive.

\item \textbf{Displaced decays:} For lifetimes in the cm–100 m range, photons originate away from the primary vertex, producing a striking, low-background signal. Identification relies on photon conversion or angular pointing. This signature is ideal for beam-dump and fixed-target experiments (SHiP~\cite{SHiP:2015vad}, FASER~\cite{FASER:2019aik}, MiniBooNE beam dump mode~\cite{MiniBooNE:2008yuf}, FACET~\cite{Cerci:2021nlb}), and can also be probed at colliders for sub-meter displacements.

\item \textbf{Delayed decays:} Even if the displacement is unresolved, delayed photon arrival can occur when the HNL travels before decaying. Observing this requires sub-ns timing. ATLAS~\cite{ATLAS:2022vhr}, CMS~\cite{CMS:2019zxa}, DUNE~\cite{Schwetz:2020xra}, and proposed setups such as FACET~\cite{Cerci:2021nlb} could be sensitive. This regime extends coverage to very long lifetimes but demands excellent timing and background rejection.
\end{itemize}

The probability that an HNL decays within a fiducial region bounded by $L_1$ and $L_2$ from the production point is~\cite{Barducci:2024nvd}
\begin{equation}
P_\text{dec}(L_1,L_2) = \exp\left(-\frac{L_1}{L_\text{dec}}\right) - \exp\left(-\frac{L_2}{L_\text{dec}}\right),
\end{equation}
with $L_{\rm dec}$ the lab-frame decay length. 
This formula applies both in long-baseline neutrino setups (T2K~\cite{T2K:2019jwa}, DUNE~\cite{{DUNE:2015lol}}, MiniBooNE~\cite{MiniBooNE:2008yuf}) and in collider or beam-dump experiments (Belle~II~\cite{BelleII:TDR}, LHC, FASER~\cite{FASER:2019aik}, SHiP~\cite{SHiP:2018xqw}). For prompt decays, $L_{\rm dec}\ll$ detector dimensions, giving $P_{\rm dec}\simeq 1$.  
The dominant radiative channel $N\to\nu\gamma$ produces a single energetic photon, forward-collimated by the HNL boost and accompanied by little additional activity. Sensitivity therefore depends on photon acceptance and reconstruction efficiency. Adequate angular and energy resolution are needed to separate signal photons from neutral-current backgrounds.  

For neutrino up-scattering production, the expected number of events is
\begin{equation}
N_{\text{event}} = N_T \int dE_r \,\frac{dR}{dE_r}\,\text{BR}(N\to\nu\gamma)\,P_{\text{dec}}\,\epsilon_{\text{geo}}\,\epsilon_{\text{rec}},
\end{equation}
where $N_T$ is the number of targets, $dR/dE_r$ is the differential rate~\cite{Shoemaker:2018vii}, and $\epsilon_{\rm geo}$, $\epsilon_{\rm rec}$ are the geometric and reconstruction efficiencies.  
For meson production, the analogous expression is~\cite{Barducci:2024nvd}
\begin{equation}
N_{\rm event}=\sum_{\cal M} N_{\cal M}\,\text{BR}({\cal M}\to N X)\,\text{BR}(N \to \nu\gamma)\,P_{\rm dec}\,\epsilon_{\rm sel},
\end{equation}
where $\epsilon_{\rm sel}$ is the photon-selection efficiency and $N_{\cal M}$ the number of mesons produced,
\begin{equation}
N_{\cal M}=
\begin{cases}
N_{\rm POT}\,f_{\cal M}, & \text{for fixed-target experiments (NA62~\cite{NA62:2017rwk,NA62:2020mcv}, HIKE\cite{HIKE:2023ext}, SHiP~\cite{SHiP:2018xqw})},\\[4pt]
\sigma_{pp}\,{\cal L}\,f_{\cal M}, & \text{for colliders (FASER2~\cite{FASER:2022hcn})}.
\end{cases}
\end{equation}
Here $N_{\rm POT}$ is the number of protons on target, $\sigma_{pp}$ the inelastic proton–proton cross section, ${\cal L}$ the integrated luminosity, and $f_{\cal M}$ the average meson multiplicity.

\subsection{Recoil Spectrum Distortions}
\label{recoil}
In neutrino scattering experiments probing the NDP, the signal is not a visible decay of HNL, but instead a distortion of the recoil energy spectrum compared to the SM prediction. 
In particular, NEES and CE$\nu$NS offer clean channels where dipole interactions add an extra contribution to the cross section. 
The characteristic feature of this contribution is an enhancement at low recoil energies, scaling approximately as $1/E_r$, which stands in sharp contrast to the SM predictions. 
Below we outline the formalism for the recoil spectra in both NEES and CE$\nu$NS channels, emphasizing how neutrino up-scattering via the dipole operator modifies the event rates measured in detectors.
\subsubsection{NEES}
The SM prediction for flavor $\alpha$ and neutrino energy $E_\nu$ is ~\cite{Vogel:1989iv}
\begin{align}
\frac{d\sigma_{\nu_\alpha e}^{\rm SM}}{dE_r}
&= \frac{2 G_F^2 m_e}{\pi}
\left[
g_L^2 + g_R^2\left( 1 - \frac{E_r}{E_\nu} \right)^2
- g_L g_R \frac{m_e E_r}{E_\nu^2}
\right],
\label{eq:nu-e-sm}
\end{align}
where $E_r$ is the electron recoil kinetic energy, $m_e$ is the electron mass, and
\begin{eqnarray}
g_L(\nu_{e}) = \frac12 + \sin^2\theta_W, ~~~ g_L(\nu_{\mu,\tau}) = -\frac12 + \sin^2\theta_W, ~~~
g_R = \sin^2\theta_W, 
\end{eqnarray}
For antineutrinos, $g_L \leftrightarrow g_R$.
The expected number of recoil events in an energy bin $i$ is then
\begin{align}
N_i &= N_e \int_{E_r^i}^{E_r^{i+1}} dE_r , \epsilon(E_r)
\int_{E_{\rm min}(E_r)}^{E_{\rm max}} dE_\nu ,
\frac{d\Phi_\nu}{dE_\nu}
\frac{d\sigma_{\nu e}^{\rm tot}}{dE_r}(E_\nu, E_r),
\end{align}
where $N_e = Z N_{\rm target}$ is the number of electrons, $\epsilon(E_r)$ the detection efficiency, and $d\Phi_\nu/dE_\nu$ the incident neutrino flux.
The minimum neutrino energy required to produce a recoil of $E_r$ is given in Eq.~(\ref{eq:Er-min}).

Different neutrino experiments implement the detector efficiency 
$\epsilon(E_r)$ in distinct ways. 
In the GEMMA experiment~\cite{Beda:2012zz}, based on a high-purity germanium detector 
at the Kalinin Nuclear Power Plant, the detection efficiency near threshold 
is determined using pulse calibrations. The effective energy threshold has 
been reported as low as $E_{\rm th} \simeq 2.8~\mathrm{keV}$, and the efficiency 
curve around this threshold is measured rather than modeled with a closed 
analytic form~\cite{Beda:2012zz,Beda:2013mta}. 
For Borexino, the large liquid-scintillator volume provides nearly 
unity efficiency above $E_r \gtrsim 150~\mathrm{keV}$, thanks to its exceptionally 
high light yield and precise calibration procedures that reconstruct electron 
recoils directly in keV$_{\rm ee}$~\cite{Borexino:2010zht}. 
In TEXONO experiment~\cite{Deniz:2009mu}, which employs a CsI(Tl) scintillating crystal array exposed 
to reactor antineutrinos, the overall efficiency in the 3--8~MeV recoil window 
is approximately constant at $\epsilon \simeq 0.77$. This value is obtained 
after applying event-selection criteria such as cosmic-ray veto, multi-hit 
rejection, and fiducial-volume cuts, with only mild residual energy dependence 
across this range~\cite{TEXONO:2006xds,Deniz:2009mu}.

Then, the total differential cross section in the presence of the NDP is simply given by
\begin{align}
\frac{d\sigma_{\nu e}^{\rm tot}}{dE_r} 
= \frac{d\sigma_{\nu_\alpha e}^{\rm SM}}{dE_r}
+ \frac{d\sigma_{\nu e}^{\rm NDP}}{dE_r}.
\end{align}
where the dipole contribution exhibits a $1/E_r$ enhancement at low recoil energy.
\subsubsection{CE$\nu$NS}
The SM prediction for the  differential cross section of \cen  with a spin-zero nucleus  $\mathcal{N}$ with $Z$ protons and $N$ neutrons  as a function of the nuclear kinetic recoil energy $E_r$  is given by \cite{Drukier:1984vhf,Barranco:2005yy,Patton:2012jr}
\begin{align}
\frac{d\sigma_{\nu_\alpha- \mathcal{N}}}{dE_{r}}(E_\nu, E_{r})=\frac{G_F^2 m_{\mathcal{N}}}{\pi}
\left( 1-\frac{m_{\mathcal{N}}E_r}{2E^2_\nu} \right) Q^2_{\alpha,{\rm SM}}, \label{crossx}
\end{align}
where $G_F$ is the Fermi constant, the index $\alpha$ denotes the neutrino flavor, $E_\nu$ is the neutrino energy,  $m_{\mathcal{N}}$ is the mass of nucleus and
$ Q_{\alpha,{\rm SM}}=[g^{\rm p}_V(\nu_\alpha) Z F_Z(|\vv{q}|^2)+g^{\rm n}_V(\nu_\alpha) N F_N(|\vv{q}|^2)]$.
For neutrino-proton coupling, $g^{\rm p}_V$ and the neutrino-neutron coupling, $g^{\rm n}_V$,  we can take  
\cite{ParticleDataGroup:2024cfk};
$g^{\rm p}_V(\nu_e) = 0.0401$, $g^{\rm n}_V(\nu_\mu) = 0.0318$, and $g^{\rm n}_V = 0.0401$.
In Eq. (\ref{crossx}), $F_Z(|\vv{q}|^2)$ and  $F_N(|\vv{q}|^2)$ are, respectively, the form factors of the proton and neutron distributions in the nucleus, respectively.
The Helm form factor is given by Eq.(\ref{eq:Helm-FF}).
The nucleon charge radius is given by $R^2_{p(n)}=\frac{3}{5} R^2_0+3 s^2$ with $R_p$ and $R_n$ are the rms radii of the proton and neutron distributions, respectively,
The size of the neutron distribution radius $R_n$ is taken to be 4.7 fm and 4.1 fm for the analyses involving CsI
and Ar, respectively \cite{Cadeddu:2020lky}.
Note that the coherence is lost for $|\vv{q}| R_{p(n)} \gtrsim 1$

The theoretical prediction of \cen event-number $N_i$ in each nuclear-recoil energy-bin $i$ is given by~\cite{Dasgupta:2021fpn}
\begin{align}
N_i = N(\mathcal{N})_D \int^{E_{r}^{i+1}}_{E^i_{r}} dE_{r} f(n_{\rm PE}(E_{r}))
\int^{E_{\rm max}}_{E_{\rm min}} dE \sum_{\nu=\nu_e,\nu_\mu,\bar{\nu}_\mu}
\frac{d\Phi_\nu}{dE}\frac{d\sigma_{\nu_\alpha- \mathcal{N}}}{dE_{r}}(E,E_{r}) \label{events}
\end{align}
where $f(n_{\rm PE})$ is the energy-dependent reconstrcution efficiency,  $E_{\rm max}=m_\mu /2 \sim 52.8$ MeV,
$N_D$ represents the number of target nuclei in the detector mass and is given by $
N_D = g_{\rm mol} \frac{m_{\rm det}N_A}{(m_{\mathcal{N}})_{\rm mol}}$
where $m_{\rm det}$ is  the detector mass, $N_A$ is the Avogadro’s number, $(m_{\mathcal{N}})_{\rm mol}$ is the molar mass of the detector molecule and $g_{\rm mol}$ is the number of atoms in a single detector molecule.
For CsI detector \cite{COHERENT:2017ipa,COHERENT:2018imc},  $m_{\rm det}=14.6 ~{\rm kg}$ and  $(m_{\rm CsI})_{\rm mol}=259.8~{\rm gram}/{\rm mol}$, and
for Ar detector \cite{COHERENT:2020iec},  $m_{\rm det}=24~{\rm kg}$ and  $(m_{\rm Ar})_{\rm mol}=39.96~{\rm gram}/{\rm mol}$.
The lower integration limit in Eq. (\ref{events}) $E_{\rm min}$ is
the minimum neutrino energy required to attain a recoil energy $E_{r}$, which is given by Eq.(\ref{eq:Er-min}).

The recostruction efficiency for CsI detector is given in terms of the detected number of photoelectrons $n_{\rm PE}$ by the function~\cite{COHERENT:2018imc}
\begin{align}
f(n_{\rm PE})=\frac{a}{1+\exp(-k(n_{\rm PE}-n_0))} \Theta(n_{\rm PE}-5),
\end{align}
where $a=0.6655^{+0.0212}_{-0.0384}, k=0.4942^{+0.0335}_{-0.0131}, n_0 =10.8507^{+0.1838}_{-0.3995}$,
and the function $\Theta(x)$ is defined as $0 (x<5), 0.5 (5\leq x <6), 1 (x\geq 6)$.
For Ar detector, we take the detector efficiency from the results in Ref. \cite{COHERENT:2020iec}.
For CsI detector, we consider the quenching factor \cite{COHERENT:2018imc},
\begin{align}
n_{\rm PE}=1.17 \left( \frac{E_{r}}{\rm keVnr}\right),
\end{align}
describing the number of photo-electrons detected by photo-multiplier tubes per keV nuclear recoil energy.
It can be used to map $n_{\rm PE}$ to the recoil energy $E_{r}$ in the analysis.
For Ar detector, the electron-equivalent recoil energy $T_{\rm ee}[{\rm keV}_{\rm ee}]$, is transformed into the nuclear recoil energy $E_{r}[{\rm keV}_{\rm nr}]$
thanks to the relation~\cite{COHERENT:2020iec}
\begin{align}
T_{ee}=f_Q(E_{r}) E_{r},
\end{align}
where $f_Q$ is the quenching factor, which is the ratio between the scintillation light emitted in the nuclear and electron recoils. It is  parameterised as $f_Q(E_{r}) =(0.246\pm 0.006~ {\rm keV}_{\rm nr}) + ((7.8\pm 0.9)\times 10^{-4})E_{r}$ up to 125 ${\rm{\rm keV}_{\rm nr}}$, and is kept constant for larger values~\cite{COHERENT:2020iec}. 
Note that the two functions $f(n_{\rm PE})$ and $f_Q(E_{r})$ play different roles in the analysis. 
The quenching factor $f_Q(E_{r})$ encodes the scintillation response of the detector medium, mapping the true nuclear recoil energy $E_{r}$ to the electron-equivalent visible energy $T_{\rm ee}$. 
In contrast, the reconstruction efficiency $f(n_{\rm PE})$ describes the probability that an event with a given number of photoelectrons $n_{\rm PE}$ is successfully triggered and recorded by the detector. 

The differential cross section for the up-scattering process, $\nu_L + \mathcal{N}\rightarrow N + \mathcal{N} $,  is given by   \cite{Magill:2018jla,Harnik:2012ni,Balantekin:2013sda, Brdar:2020quo}
\begin{eqnarray}
\frac{d\sigma }{dE_{r}}
&=&\alpha~ \mu^2_{\nu N} Z^2 F^2_N((|\vv{q}|^2)
\left[ \frac{1}{E_{r}}-\frac{1}{E_\nu}+M^2_N \frac{E_{r}-2E_\nu -m_{\mathcal{N}}}{4E^2_\nu E_{r} m_{\mathcal{N}}}+M^4_N \frac{E_{r}-m_{\mathcal{N}}}{8E^2_\nu E^2_{r} m^2_{\mathcal{N}}} \right], \nonumber \\
&& \label{em-cx}
\end{eqnarray}
where $\mu_{\nu N}$ are common for all flavors.
From Eq.(\ref{eq:Er-max}), we can obtain the maximum possible recoil energy for a given neutrino energy $E_{\nu}$ by replacing $M_T$ with $m_{\mathcal{N}}$.
In the analysis, $M_N$ is taken so that $E^{\rm max}_{r}$ becomes larger
than $E^{i}_{r}$ for a given range of $E_{\nu}$ in eq.(\ref{events}). We find that the most conservative upper limit on $M_N$ satisfying $E^{\rm max}_{r}\gtrsim E^{i}_{r}$ is about $40$ MeV, which is taken as an upper limit for the scan of $M_N$ in our analysis.
The presence of $Z^2$ scaling (as opposed to $N^2$) and the infrared divergence at low $E_{r}$ distinguish this contribution from the SM background.

\section{Experimental Searches and Constraints}
\label{sec:strategy}
Based on the experimental signatures discussed in Sec.~\ref{sec:exp-sign}, we now turn to the corresponding searches and constraints.

\emph{(i) Fixed–target accelerator experiments}, in which intense neutrino beams from meson decays enable production or up-scattering of heavy states.  
This class includes the short–baseline anomalies (LSND~\cite{LSND:2001akn}, MiniBooNE~\cite{MiniBooNE:2018esg}) as well as dedicated scattering measurements (CHARM–II~\cite{CHARM-II:1989srx}, NOMAD~\cite{NOMAD:1997pcg}, DONUT~\cite{DONUT:2001zvi}, T2K~\cite{T2K:2019bbb}).  

\emph{(ii) High–energy neutrino probes}, which extend sensitivity to the highest accessible energies.  
These include collider searches, where $e^+e^-$ or hadronic collisions produce heavy states accompanied by photons (LEP mono–photon studies~\cite{ALEPH:1993pqw,OPAL:2000puu}), and astrophysical observatories such as IceCube~\cite{IceCube:2016umi}, which detect atmospheric and astrophysical neutrinos up to the PeV scale.  

\emph{(iii) Recoil–based detectors}, where low energy neutrino scattering produces measurable recoils.  
This category covers NEES with solar and reactor fluxes (Borexino~\cite{Agostini:2017ixy}, Super–Kamiokande~\cite{Super-Kamiokande:2016yck}, GEMMA~\cite{Beda:2013mta}, TEXONO~\cite{Deniz:2009mu}, XENON1T~\cite{XENON:2020rca}) and CE$\nu$NS, first observed with accelerator neutrinos by COHERENT~\cite{COHERENT:2017ipa} and now also probed at reactors (CONUS+~\cite{CONUS:2024lnu},Dresden-II~\cite{AristizabalSierra:2022axl}) and in multi-ton dark–matter detectors pursuing solar neutrino CE$\nu$NS (XENONnT~\cite{XENON:2024ijk}, PandaX–4T~\cite{PandaX:2024muv}, LZ~\cite{LZ:2022lsv}). 

\begin{figure}[t]
\centering
\includegraphics[width=0.49\textwidth]{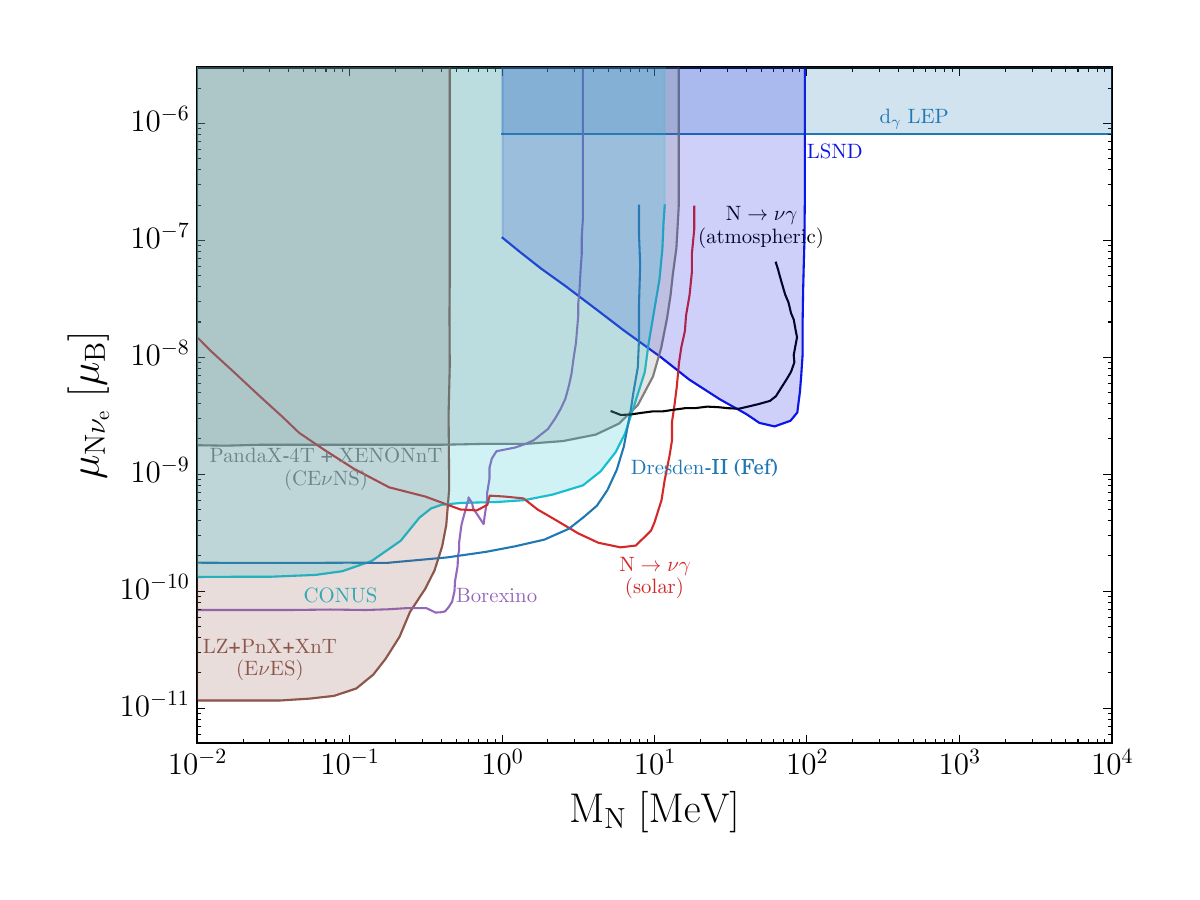}
\includegraphics[width=0.49\textwidth]{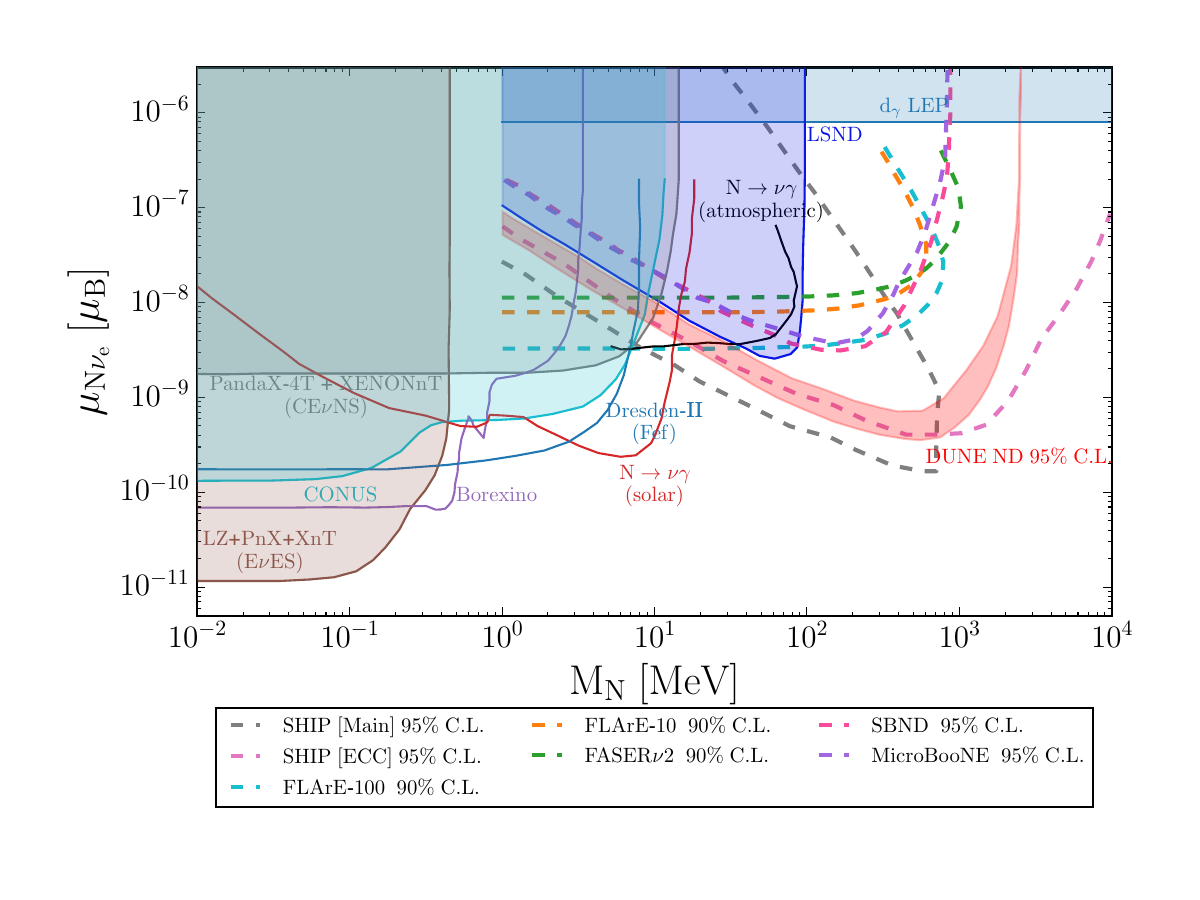}
\caption{Constraints on the $(M_N,\,\mu_{\nu_e N})$ parameter space for $\nu_e$.
\textbf{Left:} Current $90$–$95\%$ C.L. exclusions from reactor NEES (CONUS~\cite{CONUS:2020skt}, Dresden-II~\cite{AristizabalSierra:2022axl}, GEMMA~\cite{Beda:2012zz,Beda:2013mta}, TEXONO~\cite{Deniz:2009mu}), 
solar NEES searches (Borexino~\cite{Agostini:2017ixy}, SK~\cite{Super-Kamiokande:2016yck}), 
LSND~\cite{LSND:2001akn}, and dark matter detectors (XENONnT~\cite{XENON:2024ijk}, PandaX–4T~\cite{PandaX:2024muv}, LZ~\cite{LZ:2022lsv}).  
\textbf{Right:} Same as left panel, with the addition of the new CONUS+ bound (including CE$\nu$NS and NEES)~\cite{CONUS:2024lnu,DeRomeri:2025csu}, and projected sensitivities from DUNE–ND~\cite{DUNE:2015lol}, SHiP~\cite{SHiP:2015vad}, FASER$\nu$2~\cite{FASER:2022hcn}, and FLArE–10/100~\cite{Cerci:2021nlb}.
}
\label{fig:res_sterile_dipole_e}
\end{figure}

\begin{figure}[t]
\centering
\includegraphics[width=0.49\textwidth]{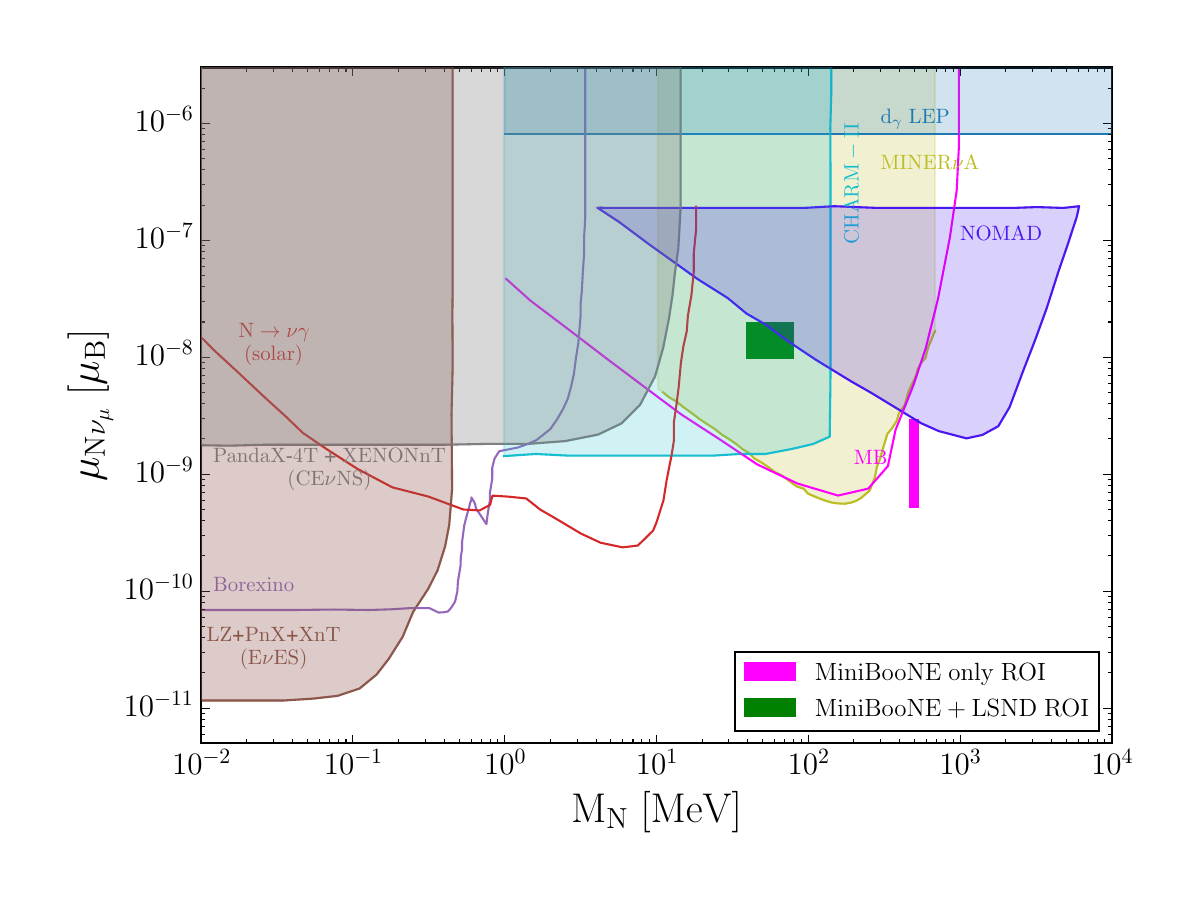}
\includegraphics[width=0.49\textwidth]{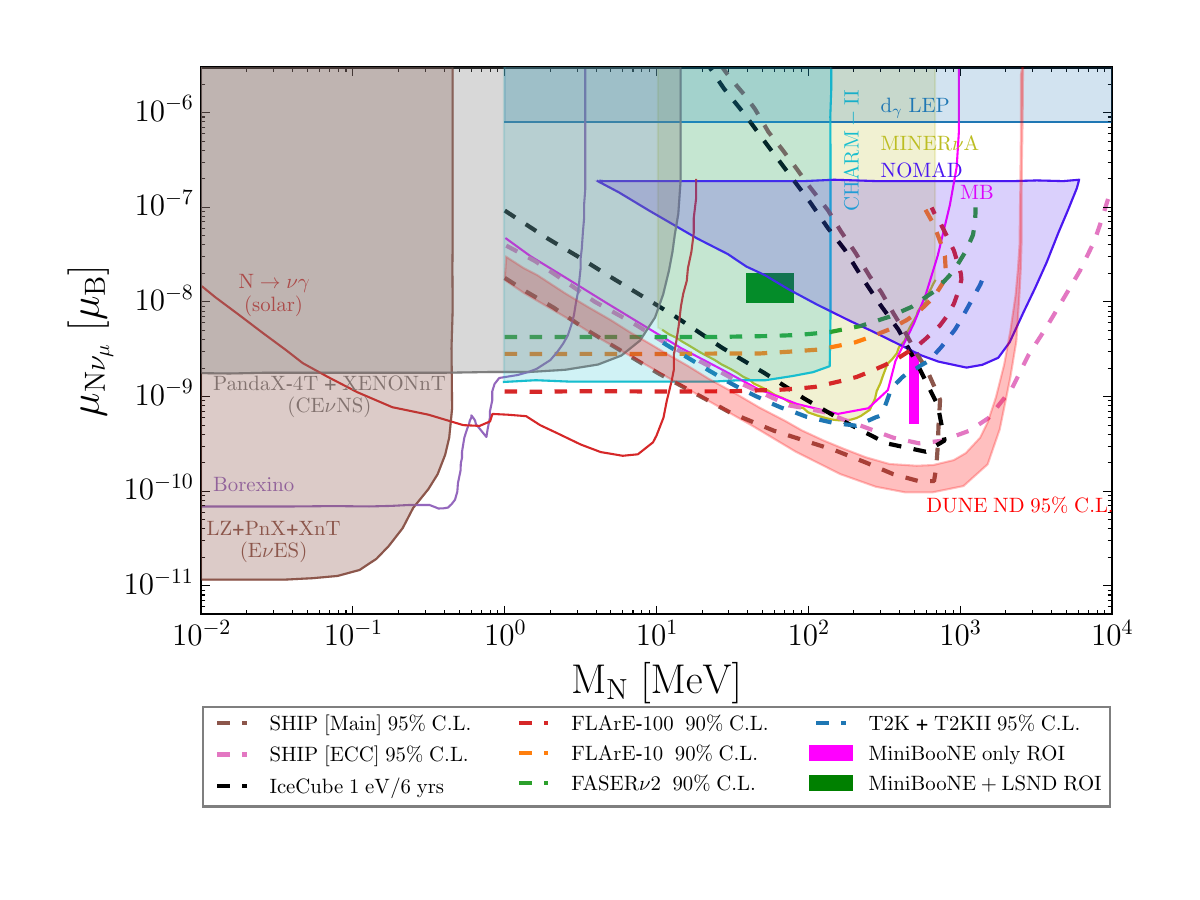}
\caption{Constraints on the $(M_N,\,\mu_{\nu_\mu N})$ parameter space for $\nu_\mu$.
\textbf{Left:} Current $90$–$95\%$ C.L. exclusions from CHARM–II~\cite{CHARM-II:1989srx}, NOMAD~\cite{NOMAD:1997pcg}, MINER$\nu$A~\cite{MINERvA:2022vmb}, Borexino, and MiniBooNE/LSND, along with CE$\nu$NS and NEES limits from Fig.~\ref{fig:res_sterile_dipole_e}. 
\textbf{Right:} Same as the left panel, with the addition of projected sensitivities from T2K+T2K–II~\cite{T2K:2019jwa}, DUNE–ND~\cite{DUNE:2015lol}, SHiP~\cite{SHiP:2018xqw}, FASER$\nu$2~\cite{FASER:2022hcn}, FLArE–10/100~\cite{Cerci:2021nlb}, and IceCube~\cite{IceCube:2016umi}. ROIs motivated by the MiniBooNE anomaly~\cite{MiniBooNE:2018esg} are also shown.}
\label{fig:res_sterile_dipole_mu}
\end{figure}

\begin{figure}[t]
\centering
\includegraphics[width=0.49\textwidth]{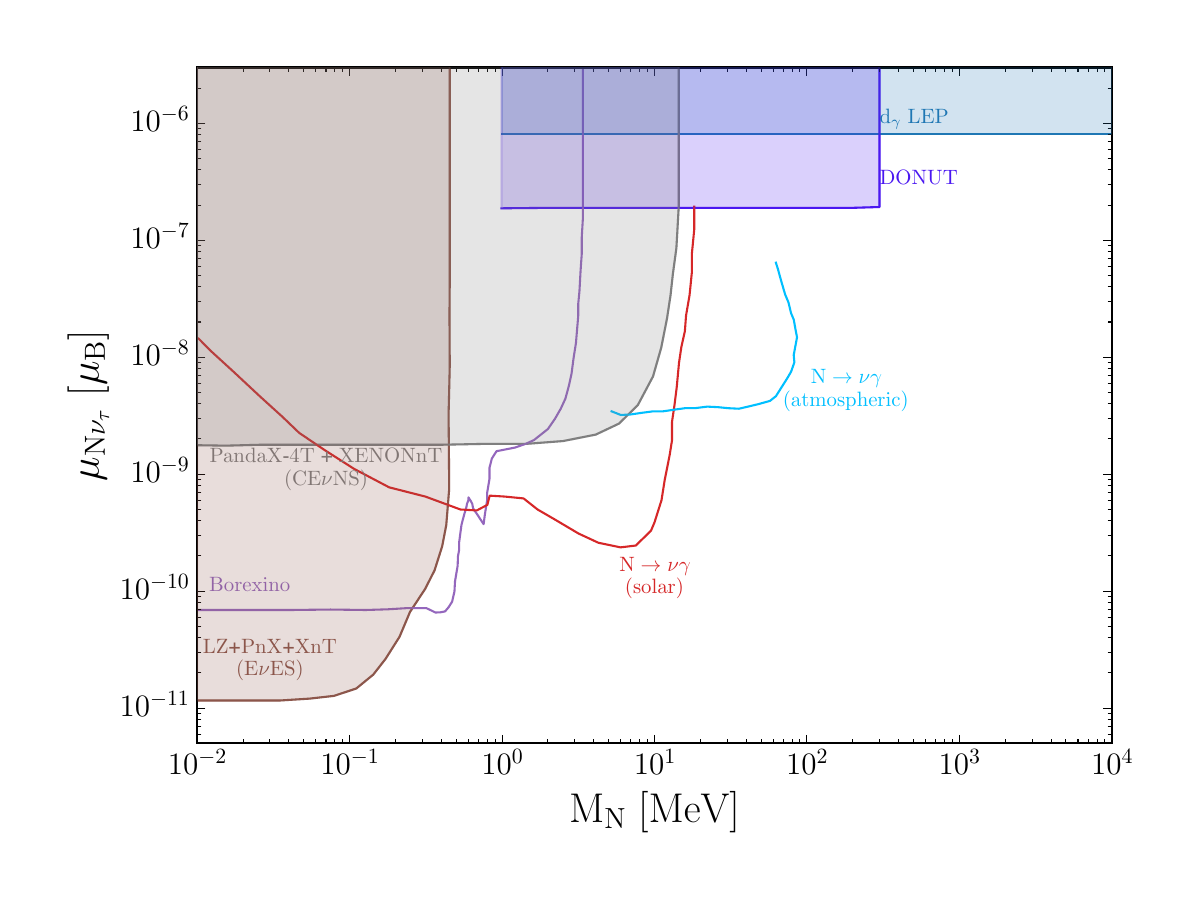}
\includegraphics[width=0.49\textwidth]{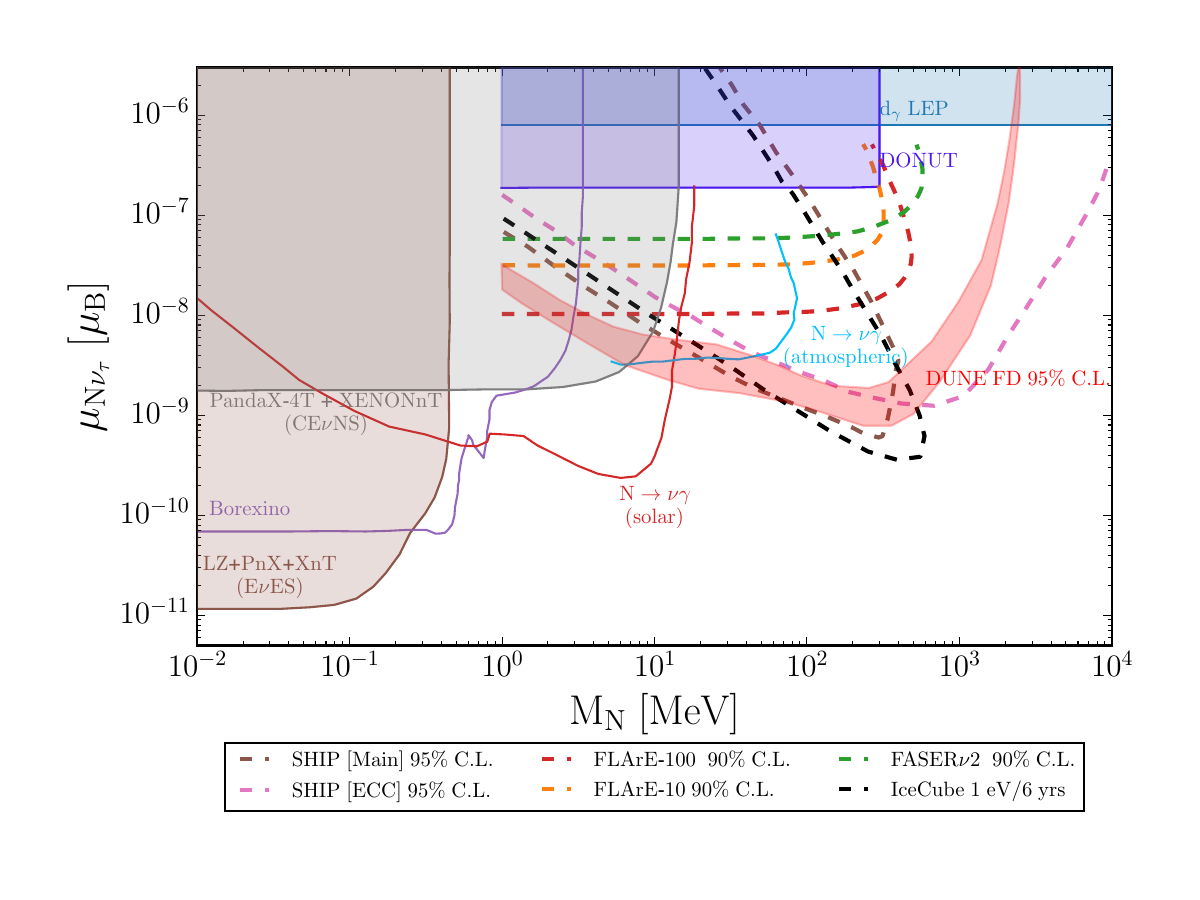}
\caption{Constraints on the $(M_N,\,\mu_{\nu_\tau N})$ parameter space for $\nu_\tau$.
\textbf{Left:} Existing $90$–$95\%$ C.L. exclusions from DONUT~\cite{DONUT:2001zvi}, solar NEES searches, reactor and CE$\nu$NS searches with xenon detectors, and radiative decay bounds from solar/atmospheric neutrinos. 
\textbf{Right:} Same as the left panel, with the addition of the projected reach of DUNE–FD~\cite{Schwetz:2020xra}, SHiP~\cite{SHiP:2018xqw}, FASER$\nu$2~\cite{FASER:2022hcn}, FLArE–10/100~\cite{Cerci:2021nlb}, and IceCube~\cite{IceCube:2016umi}, which provide sensitivity to long-lived radiative decays in the $\nu_\tau$ sector.}
\label{fig:res_sterile_dipole_tau}
\end{figure}

These complementary approaches span wide ranges in energy, baseline, and lifetime, thereby mapping out the experimentally accessible parameter space of the NDP.
Table~\ref{tab:exp_summary} summarizes the main experimental probes of the NDP, listing the neutrino source, dominant flavor sensitivity, key observables, and the accessible mass-lifetime range.
It highlights the complementarity of different approaches: solar and reactor experiments constrain the sub-MeV regime through recoil spectra, accelerator facilities probe the MeV–GeV range via up-scattering and photon signatures, and colliders and neutrino observatories extend reach to the multi-GeV and astrophysical scales.
In what follows, we present the current exclusion results and the projected sensitivities achievable at future experiments, as illustrated in Figs.~\ref{fig:res_sterile_dipole_e}–\ref{fig:res_sterile_dipole_tau}. 
Unless noted otherwise, \emph{shaded regions} or \emph{solid curves} denote existing $90$–$95\%$ C.L. exclusions, while \emph{dashed curves} indicate future sensitivities. 
When multiple analyses overlap, only the combined envelope is shown. 
Figure~\ref{fig:res_sterile_dipole_e} (electron flavor) presents results in the $(M_N,\mu_{\nu_e N})$ plane, with analogous constraints for $\nu_\mu$ and $\nu_\tau$ given in Figs.~\ref{fig:res_sterile_dipole_mu} and \ref{fig:res_sterile_dipole_tau}. 
In these figures, the left panels display current exclusions, while the right panels show the same plots with projected sensitivities overlaid as dashed curves.

\subsection{LSND and MiniBooNE anomalies}
Both LSND and MiniBooNE were short–baseline \emph{fixed–target} experiments originally designed for oscillation studies. Their anomalous $e$–like signals have been reinterpreted as possible dipole–induced upscattering followed by radiative decays, motivating early bounds on the NDP.
However, careful recasts of their data now show that the dipole explanation of these anomalies is strongly disfavored.

\textbf{[LSND experiment]:}  
The dark blue region labeled "LSND" in Fig.~\ref{fig:res_sterile_dipole_e} shows the $95\%$ C.L. exclusion obtained from LSND’s precise NEES measurement~\cite{Magill:2018jla} in the dipole–portal parameter space. Within the NDP framework, $\nu_e$ from $\mu^+$ DAR (and subleading $\nu_\mu$ from $\pi$ DIF) can upscatter on electrons via the dipole interaction, and the produced $N$ decays radiatively, yielding single-photon $e$–like Cherenkov signatures~\cite{Gninenko:2010pr,Masip:2012ke}. 
Because LSND had sensitivity to both $\nu_e$ and $\nu_\mu$ components of the DAR/DIF flux, the analysis provides complementary exclusions in both the electron– and muon–flavor dipole couplings. But, the LSND bound is shown in the electron–flavor panels, since its strongest sensitivity arises from $\nu_e$ produced in $\mu^+$ decay at rest.

\textbf{[MiniBooNE experiment]:}  
The \textbf{MiniBooNE} experiment observed an excess of low energy $e$–like events~\cite{MiniBooNE:2018esg}, motivating dipole interpretations~\cite{Gninenko:2009ks,Gninenko:2010pr,Masip:2012ke} in which $\nu_\mu$ upscatters to $N$ and  the subsequent $N\!\to\!\nu\gamma$ decay mimics an $e$–like signal.
Unlike LSND, the MiniBooNE flux was overwhelmingly dominated by $\nu_\mu$ from pion DIF, so its analysis constrains only the muon–flavor dipole coupling.
The solid magenta curve labeled "MB" in Fig.~\ref{fig:res_sterile_dipole_mu} shows a $95\%$ C.L. exclusion from binned fits to the published $E_\nu^{\rm QE}$ and $\cos\theta$ distributions~\cite{Vergani:2021tgc}, restricted to $E_\nu^{\rm QE}>475$ MeV and using a shape–only fit for $\cos\theta$ with profiled normalization~\cite{Magill:2018jla}. These bounds strongly disfavour the dipole explanation across a wide mass–coupling range. The "MiniBooNE–only" and "MiniBooNE+LSND" ROIs highlighted in Fig.~\ref{fig:res_sterile_dipole_mu} are now tightly constrained by LSND NEES and MiniBooNE anomaly–free data~\cite{Magill:2018jla}.

\subsection{Accelerator–based neutrino experiments}
\label{sec:accel-neutrino}
Beyond the anomalies, additional \emph{fixed–target} accelerator experiments set strong constraints on dipole couplings, as summarized below.

\textbf{[Existing Exclusions]:}  
In Fig.~\ref{fig:res_sterile_dipole_mu}, \textbf{CHARM–II}~\cite{CHARM-II:1989srx} and \textbf{NOMAD}~\cite{NOMAD:1997pcg} provide the leading $95\%$ C.L. bounds on the $\nu_\mu$–$N$ dipole coupling~\cite{Coloma:2017ppo,Shoemaker:2018vii}.
The cyan shaded region corresponds to the CHARM–II exclusion, derived from precision measurements of $\nu_\mu e^-\!\to\!\nu_\mu e^-$ scattering, where dipole effects enhance the low-recoil spectrum.  
The dark blue shaded region denotes the NOMAD exclusion, obtained from its dedicated single–photon search that probes dipole-induced neutrino up-scattering followed by $N\!\to\!\nu\gamma$.  
We note that these two results are complementary: CHARM–II constrains the low-recoil regime, while NOMAD probes the sub–GeV mass range via photon signatures~\cite{Coloma:2017ppo,Shoemaker:2018vii,Gninenko:1998nn,Gninenko:1998pm}.  
In addition, the \textbf{MINERvA} experiment has recently provided complementary constraints. 
Using the NuMI beam at Fermilab, MINERvA performed a search for dipole-induced single-photon signatures arising from neutrino up-scattering followed by $N \!\to\! \nu\gamma$~\cite{MINERvA:2022vmb}. 
No excess above background was observed, leading to a $95\%$ C.L. exclusion, shown as the yellow shaded regions in Fig.~\ref{fig:res_sterile_dipole_mu}, which further restricts the dipole portal parameter space in the sub-GeV mass range. 
This result is complementary to those from CHARM-II and NOMAD, providing coverage that lies between their respective sensitivity regions.
In the left panel of Fig.~\ref{fig:res_sterile_dipole_tau}, \textbf{DONUT}~\cite{DONUT:2001zvi} sets the first direct accelerator limit on $\mu_{\nu_\tau N}$ through $\nu_\tau e^-$ scattering, shown as the dark blue shaded region~\cite{Coloma:2017ppo,Shoemaker:2018vii}.

A constraint has also been derived from the \textbf{T2K ND280} (2019) dataset~\cite{T2K:2019jwa}.  
Dipole-induced upscattering $\nu A \to N A$ with subsequent $N \to \nu\gamma \to e^+e^-$ decays was searched for in the electron-like sample. 
The resulting exclusion disfavors part of the parameter space proposed as a dipole-portal explanation of the MiniBooNE anomaly, already ruling out the combined energy- and angular-distribution region at 95\% C.L.~\cite{Liu:2024cdi} 
However, the ND280 contour is not included in Fig.~\ref{fig:res_sterile_dipole_e}, since its reach is concentrated in this anomaly motivated window, while broader coverage at low and intermediate masses is already provided by LSND and by reactor/solar experiments~\cite{Liu:2024cdi}.  

\textbf{[Projected Sensitivities]:}  
 In the right panel of Fig.~\ref{fig:res_sterile_dipole_mu}, the \emph{projected} \textbf{T2K+T2K-II} sensitivity~\cite{Liu:2024cdi} is shown as a blue dashed contour.  
Because the J-PARC beam is $\nu_\mu$-dominated, the future projections are displayed in the muon-flavor panel.  
The proposed \textbf{SHiP} experiment at the CERN SPS~\cite{SHiP:2018xqw} will probe $\mu_{\nu N}$ using both the main decay volume (Main) and the emulsion cloud chamber (ECC). The projected sensitivities appear in all three right panels as dashed curves: light brown for the Main volume and pink for the ECC~\cite{Magill:2018jla}. Production from prompt mesons and dipole-induced upscattering enables lifetime–wedge coverage up to $\mathcal{O}(\mathrm{GeV})$.  
At Fermilab, \textbf{DUNE} will provide complementary reach~\cite{DUNE:2015lol}. In the right panels of Figs.~\ref{fig:res_sterile_dipole_e} and \ref{fig:res_sterile_dipole_mu}, the \emph{near detector} sensitivity is shown as a filled red contour at $95\%$ C.L., probing the neutrino up-scattering with prompt photon-like signatures from $N \!\to\! \nu\gamma$. In the right panel of Fig.~\ref{fig:res_sterile_dipole_tau}, the \emph{far detector} extends coverage to displaced $N \!\to\! \nu\gamma$ decays, probing longer lifetimes and larger $M_N$~\cite{Schwetz:2020xra}.  
Forward LHC detectors, \textbf{FASER$\nu$/$\nu2$}~\cite{FASER:2022hcn} and proposed \textbf{FLArE-10/100}~\cite{Cerci:2021nlb}, further extend the reach using TeV-scale neutrino fluxes from $pp$ collisions. Their $90\%$ C.L. sensitivities are shown by the green dashed curves for FASER$\nu$2 in each panel, and by the orange (red) dashed curves for FLArE-10(100), covering higher energies and complementing the SPS and LBNF facilities.  
We note that the anomaly-motivated \emph{MiniBooNE-only ROI} and the combined \emph{MiniBooNE+LSND ROI} highlighted in the right panel of Fig.~\ref{fig:res_sterile_dipole_mu} can be tested by the blue dashed T2K+T2K-II projection and the red DUNE-ND region, with the ND280 exclusion already disfavoring the joint energy- and angular-distribution region at $95\%$ C.L.~\cite{Liu:2024cdi}.

\subsection{Low-energy NEES experiments}
Recoil–based detectors provide leading constraints through low energy NEES, as discussed below.

\textbf{[Solar neutrino experiments]:}  
Low threshold electron recoil measurements by \textbf{Borexino}~\cite{BOREXINO:2018ohr} and \textbf{Super–Kamiokande (SK)}~\cite{Super-Kamiokande:2016yck} provide strong bounds on the $\nu_e$ dipole portal through distortions in the recoil spectrum from solar $\nu_e e \!\to\! N e$ scattering with subsequent $N \!\to\! \nu\gamma$.  
In Borexino~\cite{BOREXINO:2018ohr}, a dedicated fit to the low energy spectrum yields the strongest sensitivity for $m_N \lesssim 10$~MeV, where neutrino up-scattering is kinematically open; above this range, sensitivity falls rapidly, and the limits are dominated by the $\nu_e$ coupling. In Fig.~\ref{fig:res_sterile_dipole_e}, the purple curve labeled ``Borexino'' represents this exclusion.  
SK extends coverage to higher solar neutrino energies, and the \textbf{combined Borexino+SK analysis}~\cite{Plestid:2020vqf} further strengthens exclusions across the MeV regime by incorporating both low- and high-energy recoil data. This bound is shown by the red curve labeled ``$N\!\rightarrow\!\nu \gamma$ (solar)'' in Fig.~\ref{fig:res_sterile_dipole_e}, constraining $\mu_{\nu_e N}$ over the $m_N \sim ~{\rm MeV}$ range.

\begin{figure}[t]
\centering
\includegraphics[width=0.6\textwidth]{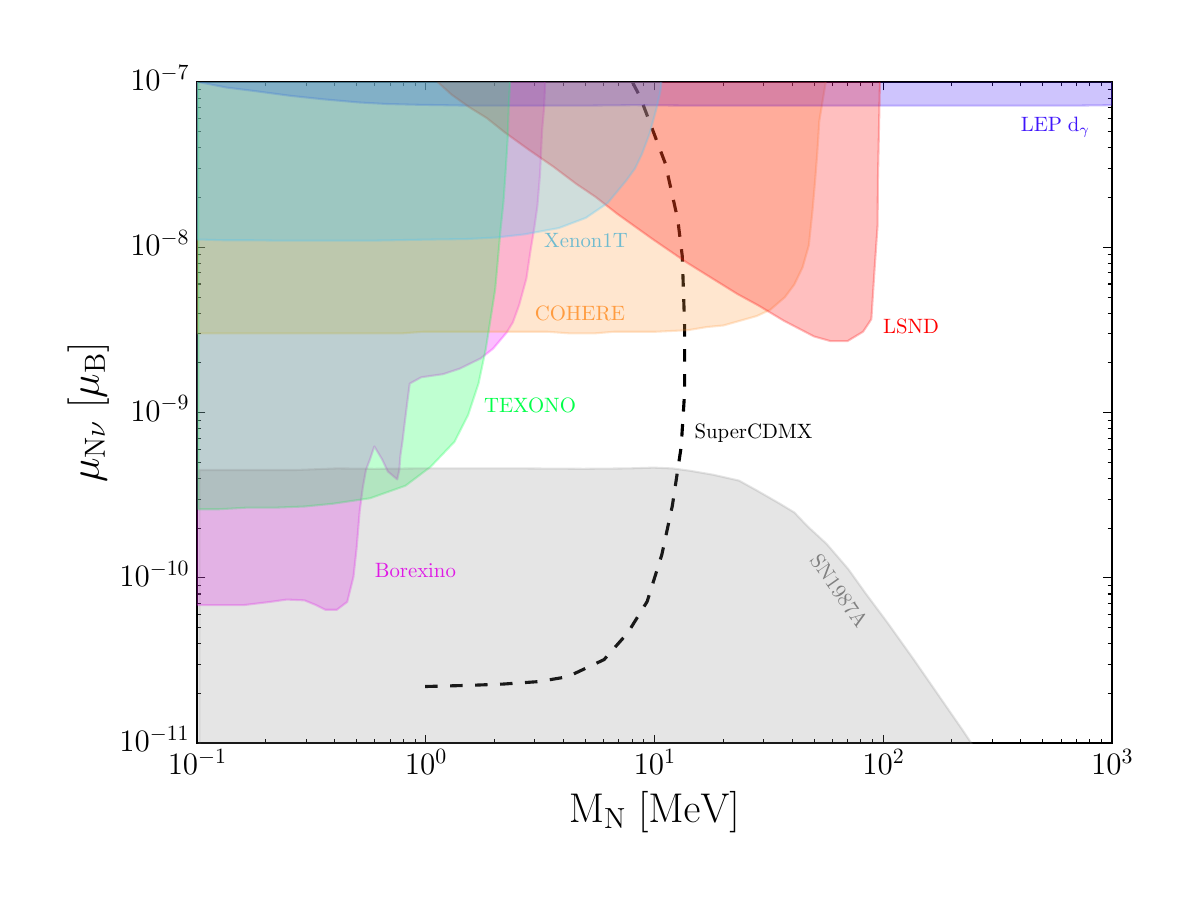}
\caption{Constraints on the $(M_N, \mu_{\nu N})$ parameter space, assuming only one dipole coupling is nonzero at a time. 
Current $90$–$95\%$ C.L. exclusions are shown from XENON1T~\cite{Shoemaker:2018vii}(blue), COHERENT~\cite{COHERENT:2017ipa}(yellow), and TEXONO~\cite{Deniz:2009mu}(green), with additional comparisons to Borexino~\cite{BOREXINO:2018ohr}(pink) and LSND~\cite{LSND:2001akn}(red). 
The projected sensitivity of SuperCDMS~\cite{SuperCDMS:2017nns} (black dashed) and the exclusion from SN1987A~\cite{Magill:2018jla} (grey) are also shown.}
\label{fig:res_sterile_dipole_x}
\end{figure}

The \textbf{XENON1T} experiment~\cite{XENON:2020rca} has derived complementary constraints on the dipole portal from the measured low energy electron recoil spectrum.  
Dipole-induced scattering would lead to an excess of events in the sub-keV to few-keV range, which is not observed. The resulting $90\%$ C.L. exclusion removes additional parameter space in the $\nu_\mu$ dipole sector, as indicated by the blue-shaded region labeled ``Xenon1T'' in Fig.~\ref{fig:res_sterile_dipole_x}~\cite{Shoemaker:2018vii}. 
The \textbf{SuperCDMS}~\cite{SuperCDMS:2017nns} experiment is projected to achieve $90\%$ C.L. sensitivity to the dipole interaction with sub-keV thresholds and ton–year exposures.  
As indicated by the black dashed curve labeled "SuperCDMS" in Fig.~\ref{fig:res_sterile_dipole_x}~\cite{Shoemaker:2018vii}, its reach extends well into the sub-MeV mass regime, fully covering the exclusion obtained from XENON1T and further lowering the bound to $\mu_{\nu N} \sim 10^{-8}\,\text{GeV}^{-1}$.
In contrast to Borexino and SK, which constrain only the electron-flavor coupling through solar $\nu_e$ scattering, the XENON1T limit and the projected SuperCDMS sensitivity are presented identically in all flavor panels under the assumption that only one dipole coupling is nonzero at a time. 

\textbf{[Reactor neutrino experiments]:}  
Reactor-based $\bar\nu_e e^-$ scattering experiments provide complementary probes of the $\nu_e$ dipole coupling.  
The \textbf{TEXONO} experiment~\cite{TEXONO:2006xds,TEXONO:2009knm}, using low-threshold germanium detectors, measures $\bar\nu_e e^-$ scattering where an NDP contribution would appear as the characteristic $1/E_e$ enhancement in the recoil spectrum.  
The resulting exclusion, shown as the green-shaded region in Fig.~\ref{fig:res_sterile_dipole_x}, constrains the dipole portal coupling in the sub-MeV to MeV mass range at $90\%$ C.L.~\cite{Bolton:2021pey,Ternes:2025lqh}.  
The \textbf{GEMMA} experiment~\cite{Beda:2013mta}, employing a high-purity germanium detector with very low recoil thresholds, is particularly sensitive to dipole-induced $\bar\nu_e e \!\to\! N e$ scattering.  
Here, the dipole interaction would enhance the event rate at sub-keV recoil energies.  
The absence of such an excess yields world leading limits in the light mass regime, corresponding to $\mu_{\nu_e N} \lesssim \text{few}\times 10^{-11}\mu_B$ at $90\%$ C.L.~\cite{Ternes:2025lqh}.

\subsection{CE$\nu$NS Experiments}
CE$\nu$NS provides a flavor blind probe of the NDP in principle, owing to the universal nuclear response, but current measurements constrain only $\mu_{\nu_e N}$ and $\mu_{\nu_\mu N}$, since available fluxes come from reactors ($\bar\nu_e$) and stopped-pion sources ($\nu_\mu, \bar\nu_\mu, \nu_e$).

\textbf{[COHERENT at the SNS]:}  
The \textbf{COHERENT} program at the Spallation Neutron Source (SNS) has provided the first and most detailed CE$\nu$NS measurements using multiple detector technologies~\cite{COHERENT:2015mry}.  
The SNS neutrino flux consists of monoenergetic $\nu_\mu$ at 29.9~MeV from $\pi^+$ decay at rest (DAR) and continuous $\nu_e$ and $\bar\nu_\mu$ spectra extending up to 52.8~MeV from $\mu^+$ decay.  
In the NDP, CE$\nu$NS cross sections receive dipole-induced enhancements that modify the recoil spectrum in a flavor dependent way: $\nu_\mu$ constraints are driven by the monoenergetic line, while $\nu_e$ and $\bar\nu_\mu$ limits follow from the DAR continuum.  
The first CE$\nu$NS observation with a CsI[Na] detector established the signal~\cite{Akimov:2017ade} and provided the initial exclusion on $\mu_{\nu N}$ under the assumption that only one dipole coupling is nonzero, as shown by the yellow-shaded region in Fig.~\ref{fig:res_sterile_dipole_e}~\cite{Dasgupta:2021fpn,Bolton:2021pey}.
Follow-up measurements with liquid argon (CENNS-10)~\cite{COHERENT:2020iec} and NaI~\cite{COHERENT:2020ybo} confirmed the effect and provided complementary constraints using different nuclear targets and systematics.  
Future germanium based detectors are expected to extend sensitivity to sub-keV recoils~\cite{COHERENT:2015mry}.  

\textbf{[Reactor CE$\nu$NS]:}  
Reactor-based experiments probe $\nu_e$ dipole couplings at much lower energies.  
The \textbf{Dresden–II} germanium detector has searched for $\bar\nu_e + \text{Ge}\!\to\!N + \text{Ge}$, applying an ON/OFF likelihood analysis that incorporates reactor flux, exposure, quenching modeled via the iron–filter method, and detector response~\cite{AristizabalSierra:2022axl}.  
The resulting 90\% C.L. exclusion rules out a significant portion of the $(M_N, \mu_{\nu_e N})$ parameter space, shown by the blue curve labeled ``Dresden-II (Fef)'' in Fig.~\ref{fig:res_sterile_dipole_e}.  
The recent \textbf{CONUS+} results~\cite{CONUS:2024lnu,DeRomeri:2025csu}, obtained with sub-keV thresholds, further strengthen these bounds.  
Under a single-flavor $\nu_e$ dipole hypothesis, the cyan solid curve in Fig.~\ref{fig:res_sterile_dipole_e} denotes the 90\% C.L. exclusion boundary, with the shaded region above it illustrating the corresponding excluded parameter space.
A characteristic feature of both Dresden–II and CONUS+ is the flat behavior at $M_N \!\ll\! E_\nu$, followed by a rise near $M_N \!\sim\! \text{few}\times 10$~MeV, reflecting the interplay of CE$\nu$NS kinematics, the reactor $\bar\nu_e$ spectrum, and nuclear form factors.  
Together, Dresden–II and CONUS+ provide the strongest reactor based CE$\nu$NS exclusions on $\mu_{\nu_e N}$.

\textbf{[Dual-phase xenon TPCs]:}  
Large dark matter detectors such as \textbf{XENONnT}~\cite{XENON:2024ijk}, \textbf{PandaX–4T}~\cite{PandaX:2024muv}, and \textbf{LZ}~\cite{LZ:2022lsv} also measure CE$\nu$NS from solar neutrinos, in addition to NEES.  
From Figs.~\ref{fig:res_sterile_dipole_e}--\ref{fig:res_sterile_dipole_tau}, the combined XENONnT+PandaX–4T CE$\nu$NS analysis yields an upper bound of $\mu_{\nu N} \lesssim 1.5 \times 10^{-10}\,\mu_B$, while the joint LZ+XENONnT+PandaX–4T NEES analysis improves this limit to $\mu_{\nu N} \lesssim 1.5 \times 10^{-11}\,\mu_B$~\cite{DeRomeri:2024hvc}.  
In the panels, these constraints are represented by the grey-shaded region (XENONnT+PandaX–4T CE$\nu$NS) and the light-brown–shaded region (LZ+XENONnT+PandaX–4T NEES), both indicating the corresponding excluded parameter space.  
In the exclusion panels, CE$\nu$NS curves exhibit an almost vertical cutoff at $M_N \!\sim\! 10$–15~MeV, while NEES bounds cut off around 3–4~MeV, both set by the kinematic endpoint of the $^8$B solar-neutrino spectrum.  
At sub-MeV masses, CE$\nu$NS limits from these multi-ton TPCs surpass Borexino’s reach, offering unprecedented sensitivity to light dipole portals.

\subsection{High–energy neutrino probes (LEP and IceCube)}
\label{sec:highenergy}
At the highest energies, both a collider (LEP) and an astrophysical neutrino observatory (IceCube) probe the NDP through high energy neutrino production and up-scattering.

\textbf{[Collider searches (LEP)]:}  
At the $Z$ pole, the \textbf{LEP} experiments searched for single-photon plus missing–energy events, 
$e^+e^-\!\to\!\gamma+\slashed{E}$~\cite{ALEPH:1993pqw}, with similar analyses later extended to higher energies at LEP2~\cite{OPAL:2000puu}. 
Within the NDP framework, this channel receives contributions from $e^+e^-\!\to\!\nu N\gamma$, where the photon can arise either from initial–state radiation or directly from the dipole vertex. 
Recasts of the LEP datasets in this context place stringent bounds on the dipole portal coupling $\mu_{\nu N}$~\cite{Magill:2018jla}. 
In the panels, the blue regions enclosed by solid blue lines (shown for $\mu_{\nu N}=9\times10^{-7}\,\mu_{\rm B}$) indicate the $95\%$ C.L. exclusion for a pure EM dipole operator.

\textbf{[IceCube]:}  
High–energy atmospheric and astrophysical neutrinos traversing the \textbf{IceCube} detector~\cite{IceCube:2016umi} can undergo dipole–induced upscattering, $\nu e\!\to\!Ne$, with the heavy state subsequently decaying radiatively, $N\!\to\!\nu\gamma$. Such processes would manifest as EM cascades, and may produce distinctive double–bang–like topologies if the upscattering and decay vertices are spatially separated~\cite{Coloma:2017ppo}. A recast of IceCube data~\cite{IceCube:2016umi} reports no excess above the expected background, leading to 95\% C.L. exclusions.  
These are shown as the black dashed lines in the right panels of Fig.~\ref{fig:res_sterile_dipole_mu} and Fig.~\ref{fig:res_sterile_dipole_tau}~\cite{Coloma:2017ppo}, corresponding to the same constraint represented by the blue dashed curve in Fig.~\ref{fig:res_sterile_dipole_x}. These bounds are most constraining at large $M_N$ and high energies, where the TeV–to–PeV neutrino flux provides maximal sensitivity. As illustrated in Ref.~\cite{Coloma:2017ppo}, the black dashed curves correspond to double–bang constraints derived from the atmospheric neutrino flux, normalized to the expectation of one visible event over six years of data taking.

LEP and IceCube provide complementary high–energy probes of the NDP: LEP delivers leading exclusions in the sub–100~GeV mass range through $e^+e^-$ collisions at the $Z$ pole, while IceCube extends sensitivity to much larger $M_N$ by exploiting atmospheric and astrophysical neutrino fluxes.


\subsection{Cosmological Constraints (BBN and CMB)}

Cosmology constrains the NDP primarily through Big Bang Nucleosynthesis (BBN) and the Cosmic Microwave Background (CMB). 
Dipole–mediated interactions such as $\nu\gamma \!\to\! N$ and $f^+f^-\!\to\! N\nu$ (with $f$ a SM fermion) can thermalize heavy neutral leptons in the early Universe at rates $\propto |\mu_{\nu N}|^2$~\cite{Magill:2018jla}.  
Once produced, their abundance and decay history determine their cosmological impact.  
During BBN ($T \sim 0.1$–1 MeV), additional relativistic degrees of freedom increase the Hubble rate and modify light–element yields, while late decays ($\tau_N \gtrsim 1$ s) inject entropy and distort the baryon–to–photon ratio and deuterium abundance~\cite{Sabti:2020yrt}.  
Consistency with observations of primordial $^4$He and deuterium therefore requires that $N$ either decouples early, decays early ($\tau_N \ll 1$ s), or never thermalizes.  
Analyses of light–element abundances, including precise D/H measurements from quasar absorption spectra~\cite{Cooke:2017cwo} and the helium fraction $Y_p$ inferred from CMB data ($Y_p = 0.243^{+0.023}_{-0.024}$ at 95\% C.L. from Planck TT,TE,EE+lowE+lensing+BAO~\cite{Planck:2018vyg}), constrain the effective number of relativistic species to $\Delta N_{\rm eff}^{\rm BBN}\lesssim 0.3$–0.4~\cite{PDG:2024BBN,Bleau:2023fsj,Planck:2018vyg}.  
For light HNL, this implies $\mu_{\nu N}\lesssim \mathcal{O}(10^{-10})\,\mu_B$, ensuring decoupling before BBN~\cite{EscuderoAbenza:2020cmq}.

\begin{figure}[t]
\centering
\includegraphics[width=0.49\textwidth]{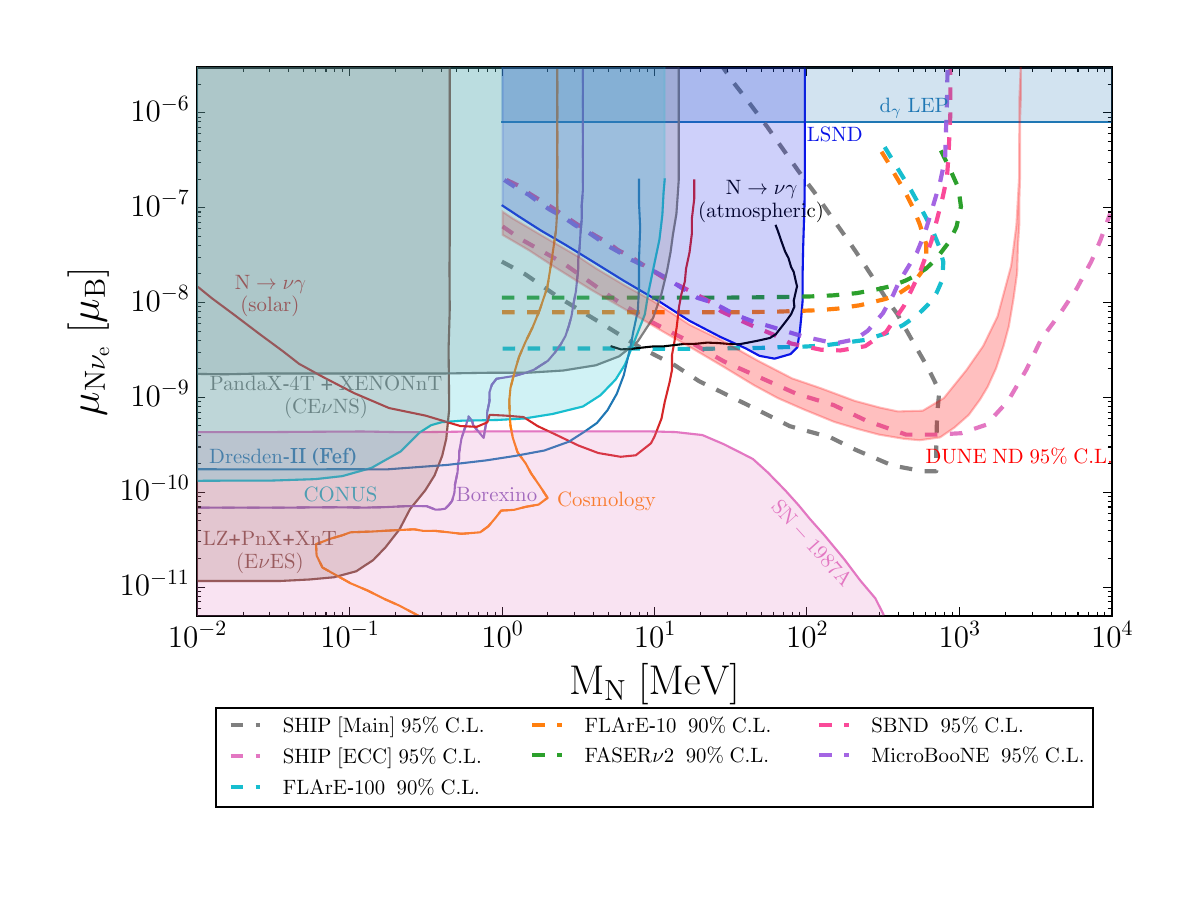}
\includegraphics[width=0.49\textwidth]{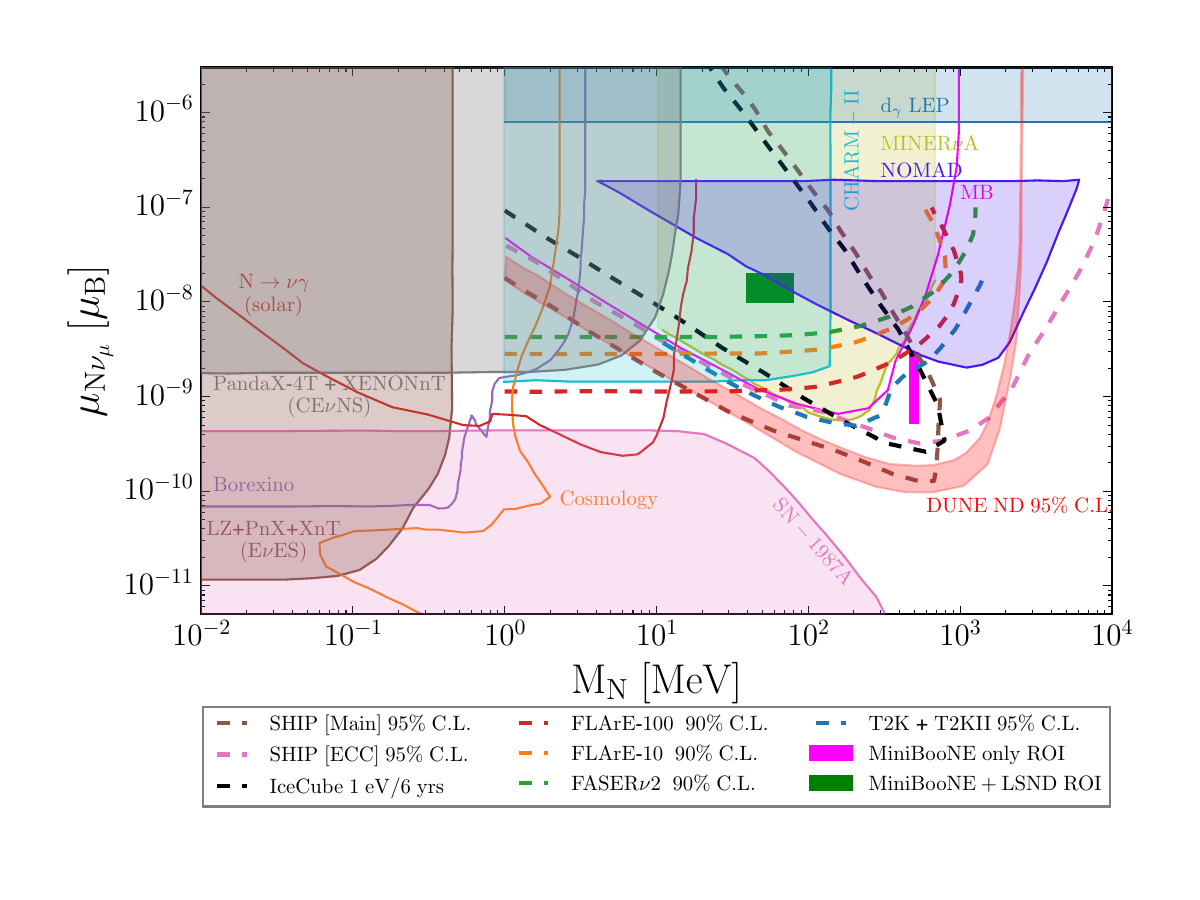}
\includegraphics[width=0.49\textwidth]{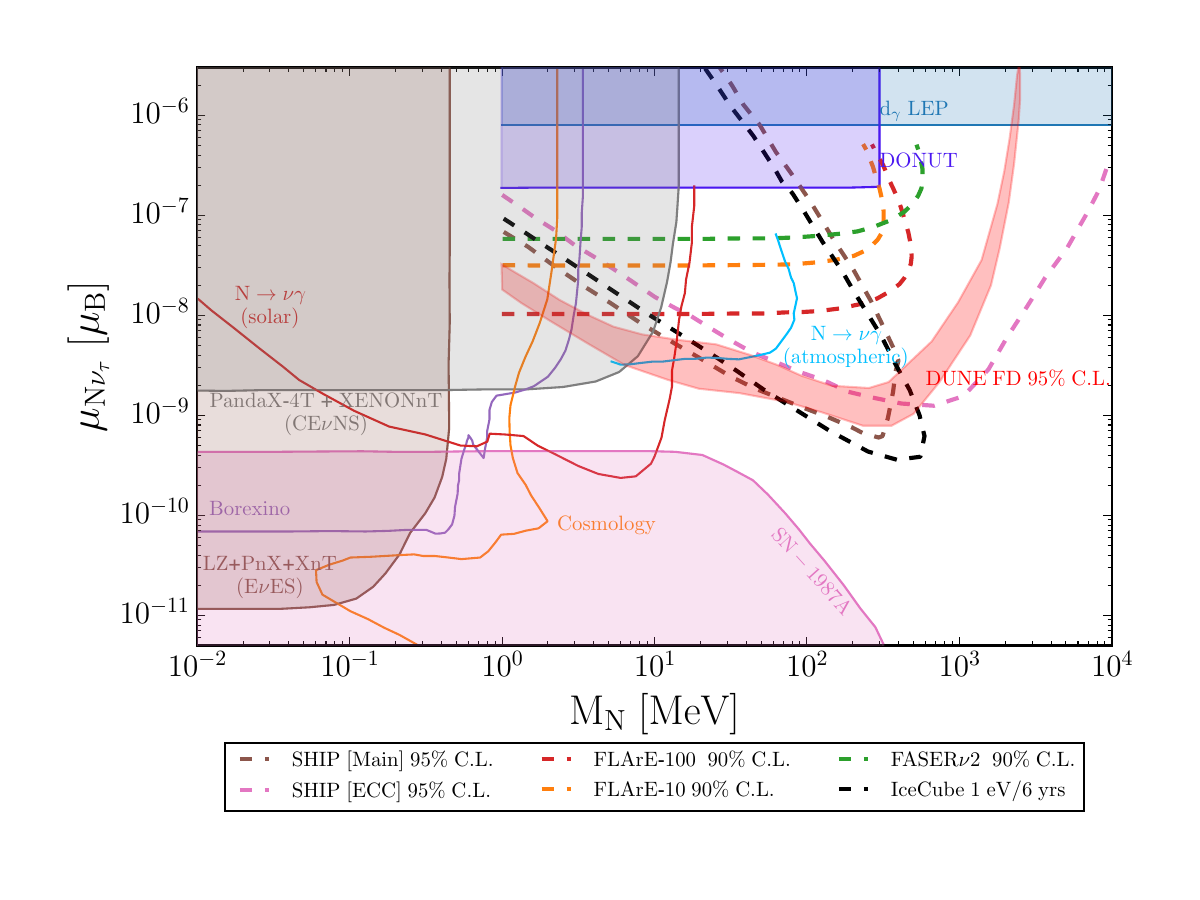}
\caption{Cosmological and astrophysical exclusions on the dipole portal parameter space, overlaid on the right panels of Figs.~\ref{fig:res_sterile_dipole_e}-\ref{fig:res_sterile_dipole_tau}.  %
Each panel shows the 95\% C.L. exclusion labeled “Cosmology,” summarizing BBN and CMB bounds on $\Delta N_{\rm eff}$ that exclude $\mu_{\nu N}\!\gtrsim\!10^{-10}\,\mu_B$ for $m_N \lesssim \mathcal{O}(10)$ MeV~\cite{DeRomeri:2024hvc}.  
Also shown is the “SN1987A” region, derived from the $\sim$10 s neutrino burst observed by Kamiokande-II~\cite{Kamiokande-II:1987idp}, IMB~\cite{Bionta:1987qt}, and Baksan~\cite{Alekseev:1987ej}, which constrains excessive energy loss from dipole-induced HNL emission~\cite{DeRomeri:2024hvc}.  }
\label{fig:res_sterile_dipole_ac}
\end{figure}

At recombination ($T \sim 0.26$ eV), the CMB provides a complementary bound on extra relativistic relics.  
The effective contribution is given by~\cite{Bleau:2023fsj,Kolb:1990vq,Mangano:2005cc,deSalas:2016ztq}
\begin{eqnarray}
\Delta N_{\rm eff}=\frac{g_N}{2}\left(\frac{T_N}{T_{\nu}}\right)^4=\frac{g_N}{2}\left(\frac{10.75}{g_{\ast,s}(T_{\rm decouple})}\right)^{4/3}
\end{eqnarray}
where we have used $g_{\ast,s}(T_{\nu})\simeq 10.75$ at the neutrino decoupling temperature $T_{\nu}$.
Tthe \textit{Planck} 2018 results combined with BAO yield $N_{\rm eff}^{\rm obs}=2.99\pm0.17$ at 68\% C.L.~\cite{Planck:2018vyg}, consistent with the SM prediction $N_{\rm eff}^{\rm SM}=3.045$~\cite{EscuderoAbenza:2020cmq}, corresponding to $\Delta N_{\rm eff}\lesssim 0.3$ at 95\% C.L.  
%
Taken together, BBN and CMB require $\mu_{\nu N}\lesssim \mathcal{O}(10^{-10})\,\mu_B$ for $m_N$ up to a few–tens of MeV, ensuring early decoupling or rapid decay.  
The exclusion labeled “Cosmology” in Fig.~\ref{fig:res_sterile_dipole_ac} summarizes the envelope of constraints derived from BBN and CMB observations~\cite{DeRomeri:2024hvc}

\subsection{Astrophysical Constraints from SN1987A}
\label{sec:astro}

Core–collapse supernovae probe the dipole portal through energy–loss arguments.  
In SN1987A, the proto–neutron star reached $T \sim 30$ MeV and nuclear densities $\rho \sim 3\times10^{14}\,\mathrm{g/cm^3}$, enabling prolific $N$ production through NEES and plasmon decay~\cite{Ayala:1998qz,Raffelt:1996wa,Magill:2018jla}.  %
If HNL couples too weakly, it free-streams out of the star, accelerating cooling beyond the observed $\sim 10$ s neutrino burst~\cite{Lattimer:1988mf,Ayala:1998qz}.  
If instead it couples too strongly, its mean free path shrinks and HNL becomes trapped, thermalizing with the plasma and decoupling from a new neutrinosphere~\cite{Caputo:2021rux,Chauhan:2024nfa}.  
The observed burst duration in \textbf{Kamiokande-II}~\cite{Kamiokande-II:1987idp}, \textbf{IMB}~\cite{Bionta:1987qt}, and \textbf{Baksan}~\cite{Alekseev:1987ej} matched expectations from SM neutrino diffusion, leaving little room for additional exotic losses.  

Imposing the Raffelt criterion~\cite{Raffelt:1996wa} on the exotic emissivity yields an approximate upper bound for light HNLs ($m_N \lesssim 100$ MeV):  
\[
\mu_{\nu N} \lesssim (1\text{--}3)\times 10^{-9}\,\mu_B.
\]  
At larger couplings, the free–streaming bound no longer applies; this defines a trapping window around  
\[
10^{-8}\,\mu_B \lesssim \mu_{\nu N} \lesssim 10^{-6}\,\mu_B,
\]  
where constraints weaken~\cite{Magill:2018jla}.  
The shaded “SN1987A” regions in Fig.~\ref{fig:res_sterile_dipole_ac} summarize this interplay of free–streaming and trapping, excluding a broad band of parameter space for $m_N \lesssim 100$ MeV that is otherwise unconstrained by laboratory data~\cite{DeRomeri:2024hvc}.

\section{Conclusion and Outlook}
\label{sec:conlcusion}
The NDP represents a minimal but powerful extension of the SM neutrino sector. 
Through a single higher-dimensional operator, it links active neutrinos to HNLs and the EM field, generating a wide spectrum of novel phenomena: radiative decays, single-photon and displaced-photon signatures, and characteristic distortions in electron and nuclear recoil spectra. 
Because the same operator governs production, decay, and scattering, the NDP provides a unifying framework in which laboratory, astrophysical, and cosmological probes all converge.

Over the past decade, an impressive range of experiments has constrained this framework. 
Short-baseline anomalies at LSND and MiniBooNE motivated early dipole-portal interpretations, now largely excluded by accelerator scattering data from CHARM–II, NOMAD, MINERvA, and T2K ND280. 
Recoil-based measurements such as Borexino, SK, Dresden–II, and CONUS+ probe the low-mass regime, while multi-ton xenon and argon TPCs have pushed sensitivities to unprecedented levels. 
At the highest energies, LEP and IceCube provide complementary exclusions, while astrophysical and cosmological arguments from SN1987A, BBN, and the CMB close off large regions of parameter space. 
Together, these inputs have already constrained $\mu_{\nu N}$ down to $\sim 10^{-11}\,\mu_B$ for HNL masses below $\mathcal{O}(10)$ MeV.

Several forthcoming developments are expected to define the next decade of exploration.
Next-generation long-baseline experiments such as DUNE and Hyper-Kamiokande will combine near and far detector capabilities to probe both recoil distortions and displaced radiative decays. 
High energy frontier experiments including SHiP, FASER$\nu$2, and FLArE will extend coverage into the multi-GeV regime, complementing astrophysical probes. 
At lower energies, Dresden–II, CONUS+, and upcoming sub-keV threshold detectors (e.g., SuperCDMS) will further test the sub-MeV mass range. 
Dark matter detectors such as XENONnT, PandaX-4T, and LZ, though designed for other purposes, will continue to play a leading role in probing light sterile states via CE$\nu$NS and NEES. 
On the cosmological and astrophysical side, improved BBN and CMB analyses, together with the next galactic supernova, promise to refine bounds on the dipole portal far beyond current laboratory reach.

The overall picture is one of complementarity. 
Accelerator programs are closing gaps in the sub-GeV to multi-GeV window; reactor and solar experiments dominate the sub-MeV regime; and astrophysical and cosmological probes guarantee coverage across and beyond laboratory sensitivities. 
This multi-pronged approach ensures that the most phenomenologically viable regions of parameter space will be either robustly excluded or tested with discovery potential in the near future.

In conclusion, the NDP provides a rare example of a simple, predictive, and highly testable framework that bridges energy, intensity, astrophysical, and cosmological frontiers. 
Its exploration constrains HNL scenarios, illuminates possible connections to hidden sectors, and sharpens our understanding of neutrino EM properties. 
The next generation of experiments and observations will determine whether the NDP remains a powerful constraint on new physics models, or emerges as the first confirmed window into dynamics beyond the SM.

%
%
%
\section*{Acknowledgments}
This work was supported by the Research Program funded by the Seoul National University of Science and Technology.


\appendix
\section{Formulae for Decay Rate}
\subsection {Two-body Decay}
For two-body decay, the decay rate is calculated by 
\begin{eqnarray}
\Gamma_{fi}=\frac{p^{\ast}}{32\pi^2 m^2_a} \int |{\cal M}_{fi}|^2 d\Omega,
\end{eqnarray}
where $p^{\ast}=\frac{1}{2m_a}\sqrt{[m^2_a-(m_1+m_2)^2][m_a^2-(m_1-m_2)^2]}$.

The HNL decays into a neutrino and a gamma via the NDP interaction, whose decay rate is given by 
\be
\Gamma_{N\to \nu\gamma}=\frac{|\mu_{\nu N}|^2M_N^3}{4\pi},
\label{decay-rate}
\ee
Using the decay rate, we can calculate  the decay length and lifetime of $N$ .
For an HNL energy of $E_N=1\GeV\gg M_N$, the decay length ($L_{\rm dec}$) and lifetime of $N$ ($\tau_N$) in lab-framework scale as
\begin{align}\begin{split}
L_\text{dec}&= c\tau\beta\gamma\approx 2500\text{m}\left(\frac{10\MeV}{M_N}\right)^4\left(\frac{10^{-6}\GeV^{-1}}{d}\right)^2, \\
\tau_N&=\tau\gamma=8.13\times 10^{-4}\text{s}\left(\frac{10\MeV}{M_N}\right)^4\left(\frac{10^{-6}\GeV^{-1}}{d}\right)^2, 
\label{eq:ldecaytdecay}
\end{split}\end{align}
where $\tau$ is the proper lifetime of $N$, and $\gamma, \beta$ its Lorentz boost factors. 
This turns out to be a very convenient length scale for beam dump experiments, if $M_N$ and $\mu_{\nu N}$ have the fiducial values suggested above.

\subsection{Three-body Decay}
We start from the total decay width for a three-body process:
\[
X(p) \to a(p_1) + b(p_2) + c(p_3),
\]
given by:
\begin{equation}
\Gamma = \frac{1}{(2\pi)^3} \frac{1}{32 m_X^3} \int |\mathcal{M}|^2 \, dm_{ab}^2 \, dm_{bc}^2.
\tag{5}
\end{equation}

Our goal is to express the differential decay rate with respect to $m_{ab}^2$ by integrating over $m_{bc}^2$:
\begin{equation}
\frac{d\Gamma}{dm_{ab}^2} = \frac{1}{(2\pi)^3} \frac{1}{32 m_X^3} \int |\mathcal{M}(m_{ab}^2, m_{bc}^2)|^2 \, dm_{bc}^2.
\tag{6}
\end{equation}

This form holds when the squared amplitude depends on two Mandelstam-like variables, $m_{ab}^2 = (p_1 + p_2)^2$ and $m_{bc}^2 = (p_2 + p_3)^2$.

\subsubsection{Integration Limits for $m_{bc}^2$}
For fixed $m_{ab}^2$, the kinematically allowed range for $m_{bc}^2$ is given by:
\[
(m_{bc}^2)_{\text{min}} \le m_{bc}^2 \le (m_{bc}^2)_{\text{max}},
\]
with the limits determined by solving the energy-momentum constraints and triangle inequalities. These can be expressed using the Källén function~\cite{Kallen}:
\begin{eqnarray}
\lambda(a,b,c) = a^2 + b^2 + c^2 - 2ab - 2ac - 2bc.
\label{kallen}
\end{eqnarray}

\subsubsection{Loop/Phase-Scape Function}
The dimensionless loop/phase-space function for three-body decay ($\pi^+\rightarrow \mu^+\gamma N$) is typically given by ~\cite{Bondarenko:2018ptm}:
\begin{eqnarray}
I(r_\mu, r_N) = \int_{r_\mu^2}^{(1 - r_N)^2} \! \mathrm{d}x \; \frac{(x - r_\mu^2)}{x^3} \, \sqrt{\lambda(1, x, r_N^2)} \, F(x),
\label{eq:loop-func}
\end{eqnarray}
where  \(x = E_\gamma / m_\pi\) is the rescaled photon energy,  \(r_\mu = m_\mu / m_\pi\), \(r_N = m_N / m_\pi\),
\(\lambda(a, b, c) \)is the Källén function given by Eq.(\ref{kallen}), arising from phase space,
\(F(x)\) is a loop integrand factor (from the virtual neutrino propagator and dipole form), which typically behaves like \(x\) or \(1/x\) depending on the operator structure.

\subsection{Four-body Decay}
\label{sec:4-body}
The differential full four-body decay rate is given by:
\begin{equation}
\dd{\Gamma} = \frac{1}{2 m_M} \left| \mathcal{M} \right|^2 \dd{\Phi}_4,
\end{equation}
where the four-body phase space is:
\begin{equation}
\dd{\Phi}_4 = (2\pi)^4 \delta^{(4)}(q - p_\mu - p_N - p_+ - p_-) \prod_{i=1}^4 \frac{\dd^3 p_i}{(2\pi)^3 2 E_i}.
\end{equation}

\subsubsection{Example: $\pi^+ \to \ell^+ \gamma \nu$}
Define a dimensionless photon energy variable:
\[
x = \frac{2 E_\gamma}{m_\pi}, \quad 0 \le x \le 1 - r_\ell^2, \quad r_\ell = \frac{m_\ell}{m_\pi}.
\]

The squared matrix element $|\mathcal{M}|^2$ typically depends on $x$ and can be integrated as:
\[
\frac{d\Gamma}{dx} = \frac{1}{(2\pi)^3} \frac{1}{32 m_\pi} \int |\mathcal{M}(x)|^2 \, d\cos\theta,
\]
where $\theta$ is the angle between $\ell^+$ and $\gamma$ in the pion rest frame. The differential rate may involve functions like $\lambda^{1/2}(1,x,r_\nu^2)$ and dipole factors such as $F(x) \sim 1/x$ or $x$.
%
%

\begin{sidewaystable}[htbp] 
\centering
\renewcommand{\arraystretch}{1.3}
\tiny
\begin{tabularx}{\linewidth}{l l l l l} 
\toprule
\textbf{Experiment} & \textbf{Neutrino source} & \textbf{Flavor sensitivity} & \textbf{Observable / strategy} & \textbf{$M_N$ reach \& lifetime} \\
\midrule
Borexino & Solar ($pp$, $^7$Be, $^8$B) & Mainly $\nu_e$ (CC-enhanced) & $e$ recoil spectrum distortion & $\lesssim 10$ MeV; prompt \\
Borexino+SK & Solar ($^8$B$+$ higher-$E$) & $\nu_e$ dominated & Low+high-$E$ recoil spectra & $\lesssim 15$ MeV; prompt \\
Super-K (atm. $\nu$) & Atmospheric (GeV) & $\nu_e$, $\nu_\mu$, $\nu_\tau$ via osc. & Up-scattering + $N\to\nu\gamma$ & 0.1–10 GeV; displaced or prompt \\
IceCube & Atm /astro (TeV–PeV) & All flavors (oscillated) & $\gamma/e$-like events from $N\to\nu\gamma$ & $\gtrsim$ GeV–TeV; long-lived \\
\midrule
XnT/ PnX / LZ & Solar ($pp$, $^8$B) & Flavor blind (CE$\nu$NS) & Electron recoil + CE$\nu$NS& $\lesssim 10$–15 MeV; prompt \\
SuperCDMS & Solar (low-$E$) & Mainly $\nu_e$, flavor blind & Sub-keV electron recoils & Sub-MeV–few MeV; prompt \\
Dresden-II & Reactor $\bar\nu_e$ & $\nu_e$ only & CE$\nu$NS upscattering on Ge & $\lesssim 10$ MeV; prompt \\
COHERENT & SNS (29.9 MeV $\nu_{(\mu,e)}$, $\bar\nu_\mu$) & $\nu_\mu,\nu_e,\bar\nu_\mu$ & CE$\nu$NS + recoil spectrum & $\lesssim 50$ MeV; prompt \\
\midrule
LSND & Accelerator (30 MeV $\nu_e$) & $\nu_e$ & $\nu_e$–$e$ ES & $\lesssim 50$ MeV; prompt \\
MiniBooNE & Booster $\nu_\mu$ beam (GeV) & $\nu_\mu$, $\nu_e$ contamination & $e$-like events from $N\to\nu\gamma$ & $\sim$0.1– GeV prompt+upstream \\
CHARM-II & SPS $\nu_\mu$ & $\nu_\mu$ & $\nu_\mu$–$e$ ES & $\sim$ GeV scale; prompt \\
NOMAD & SPS $\nu_\mu$ & $\nu_\mu$ & $\nu_\mu$–$e$ ES, mono-$\gamma$ search & GeV scale; prompt \\
DUNE ND & LBNF beam (GeV) & $\nu_\mu$ dominated, all flavors & $\nu e \to N e$ then $N\to\nu\gamma$ & 0.1–GeV; prompt \\
DUNE FD & LBNF beam (GeV) & Mixed flux after osc. & Displaced $N\to\nu\gamma$; long-lived \\
SHiP ECC & SPS (400 GeV $p$ beam) & $\nu_\mu$ dominated & High-res. $N\to\nu\gamma$ & 0.1–GeV; short-lived \\
SHiP Main & SPS beam dump & $\nu_\mu$ dominated & mono-$\gamma$ from $N\to\nu\gamma$ & 0.1–GeV; long-lived \\
\midrule
LEP (91 GeV) & $e^+e^-$ at $Z$-pole & Flavor independent  & mono-$\gamma$ + $\not\!\!E$ & $\sim 10$–100 GeV; prompt \\
\bottomrule
    Table 1. Classification of experimental bounds on the NDP.
\end{tabularx}
\caption{Classification of experimental bounds on the NDP.}
\label{tab:exp_summary}
\end{sidewaystable}

%
%

\end{document}